\documentclass[10pt, final, twocolumn, twoside, romanappendices]{IEEEtran}
\usepackage{xcolor}
\usepackage{mathtools}
\usepackage{epsfig,makeidx,color}
\usepackage{amsmath,amssymb,bbm,enumitem}
\usepackage{cite,graphicx,lipsum}
\usepackage[switch,pagewise]{lineno}
\usepackage{hyperref}
\usepackage[caption=false,font=footnotesize]{subfig}
\usepackage{algorithm}
\usepackage{algpseudocode}
\usepackage{acronym}
\usepackage{array}    
\usepackage{booktabs}
\hypersetup{
        colorlinks = true,
        citecolor=red,
}
\usepackage{tikz}

\def\cH{{\cal H}}

\def\cG{{\cal G}}

\DeclareMathOperator*{\argmax}{\arg\!\max}

\def\be{ \begin{equation} }
\def\ee{ \end{equation} }
\def\bea{ \begin{eqnarray} }
\def\eea{ \end{eqnarray} }

\def\b0{{\bf 0}}

\def\cC{{\cal C}}

\def\cD{{\cal D}}
\def\cF{{\cal F}}

\def\cI{{\cal I}}

\def\cW{{\cal W}}

\acrodef{TX}{transmitter}
\acrodef{RX}{receiver}
\acrodef{BEC}{bit erasure channel}
\acrodef{MSE}{mean squared error}
\acrodef{LPIPS}{learned perceptual image patch similarity}
\acrodef{AIGC}{AI-generated content}
\acrodef{ML}{machine learning}
\acrodef{NLP}{natural language processing}
\acrodef{LLM}{large language model}
\acrodef{CLIP}{contrastive language-image pre-training}
\acrodef{SemCom}{semantic communication}
\acrodef{SemPA-Look}{lookahead-based semantic packet aggregation}
\acrodef{SemRT}{semantic repeated transmission}
\acrodef{SMART}{semantic packet aggregation and repeated transmission}
\acrodef{GA}{genetic algorithm}
\acrodef{OOV}{out-of-vocabulary}
\acrodef{PA}{packet aggregation}
\acrodef{TC}{token communication}
\acrodef{URLLC}{ultra-reliable low-latency communication}
\acrodef{ATS}{average token similarity}
\acrodef{AIGC}{AI-generated content}
\acrodef{SemPA-GBeam}{Semantic PA with genetic beam search}
\acrodefplural{TSS}{top semantic scores}
\acrodefplural{RSS}{residual semantic scores}
\acrodef{TSS}{top semantic score}
\acrodef{RSS}{residual semantic score}

\ifCLASSOPTIONonecolumn
\else

\fi

\usepackage{geometry}
\graphicspath{ {./images/} }
\usepackage{graphicx}
\begin{document}



\title{Low-Complexity Semantic Packet Aggregation for Token Communication via Lookahead Search}



\author{
   \vspace{0.0cm}
    Seunghun~Lee,~\IEEEmembership{Graduate Student Member,~IEEE},
    Jihong~Park,~\IEEEmembership{Senior~Member,~IEEE},\\
    Jinho~Choi,~\IEEEmembership{Fellow,~IEEE},
    and Hyuncheol~Park,~\IEEEmembership{Senior~Member,~IEEE}

     
    \thanks{
        S. Lee and H. Park are with
        the School of Electrical Engineering,
        Korea Advanced Institute of Science and Technology,
        Daejeon 34141,
        Republic of Korea
        (e-mail: {seunghun21@kaist.ac.kr; hcpark@kaist.ac.kr}).

        J. Park is with
        ISTD Pillar, Singapore University of Technology and Design (SUTD),
        8 Somapah Rd, Singapore 487372, 
        Singapore
        (e-mail: {jihong\_park@sutd.edu.sg}).

        J. Choi is with the School of Electrical and Mechanical Engineering, 
        The University of Adelaide,
        Adelaide SA 5005,
        Australia
        (e-mail: {jinho.choi@adelaide.edu.au}).
    }
    \thanks{
This work was partly supported by the Institute of Information \& Communications Technology Planning \& Evaluation(IITP)-ITRC(Information Technology Research Center) grant funded by the Korea government(MSIT)(IITP-2025-RS-2023-00259991) (34\%) and in part by SUTD Kickstarter Initiative (SKI 2021\_06\_08) (33\%), and in part by the National Research Foundation, Singapore, and the Infocomm Media Development Authority under its Future Communications Research \& Development Programme (33\%).
     (\textit{Corresponding author: H. Park})
     }
\vspace{-0.8cm}
}

\maketitle

\begin{abstract} 
Tokens are fundamental processing units of generative AI (GenAI) and large language models (LLMs), and token communication (TC) is essential for enabling remote AI-generate content (AIGC) and wireless LLM applications. Unlike traditional bits, each of which is independently treated, the semantics of each token depends on its surrounding context tokens. This inter-token dependency makes TC vulnerable to outage channels, where the loss of a single token can significantly distort the original message semantics. Motivated by this, this paper focuses on optimizing token packetization to maximize the average token similarity (ATS) between the original and received token messages under outage channels. Due to inter-token dependency, this token grouping problem is combinatorial, with complexity growing exponentially with message length. To address this, we propose a novel framework of semantic packet aggregation with lookahead search (SemPA-Look), built on two core ideas. First, it introduces the residual semantic score (RSS) as a token-level surrogate for the message-level ATS, allowing robust semantic preservation even when a certain token packet is lost. Second, instead of full search, SemPA-Look applies a lookahead search-inspired algorithm that samples intra-packet token candidates without replacement (fixed depth), conditioned on inter-packet token candidates sampled with replacement (fixed width), thereby achieving linear complexity. Experiments on a remote AIGC task with the MS-COCO dataset (text captioned images) demonstrate that SemPA-Look achieves high ATS and LPIPS scores comparable to exhaustive search, while reducing computational complexity by up to 40$\times$. Compared to other linear-complexity algorithms such as the genetic algorithm (GA), SemPA-Look achieves 10$\times$ lower complexity, demonstrating its practicality for remote AIGC and other TC applications.

\end{abstract}

\begin{IEEEkeywords}
Semantic Packet Aggregation, Token Communication, Semantic Communication, Wireless AIGC.
\end{IEEEkeywords}

\ifCLASSOPTIONonecolumn
\baselineskip 28pt
\fi
\section{Introduction}

Advances in semantic communication have shifted the focus from bit‑level reliability to preserving the meaning of transmitted messages, which is essential for supporting emerging applications that convey critical semantic information through contextual processing and generative models \cite{Deniz23,Christina25}. Recent breakthroughs in large language models (LLMs) and generative AI (GenAI) have given rise to applications such as \ac{AIGC} \cite{Minrui24} and wireless LLMs \cite{oh2025HLM}, all of which rely on communicating text‑based tokens—a paradigm known as \ac{TC} \cite{qiao2025tc}. Since these tokens encode the semantics of underlying raw data (e.g., images synthesized by \ac{AIGC}, even the loss of a few tokens can drastically alter the intended meaning in context, severely degrading the target task’s effectiveness. This motivates us to revisit packetization strategies for \ac{TC} with the focus of preserving token-level semantics.

In \ac{TC}, information is represented as sequences of tokens, commonly processed using transformer‑based architectures \cite{qiao2025tc,Xia25,Liang25}, and text tokens in particular offer rich semantic content, well‑defined linguistic structure. 
Recent studies on text‑based communication have explored LLM‑based source and channel coding \cite{Nam2023LanguageOrientedCW}, error correction techniques \cite{Guo22,Guo24}, and multimodal data transmission \cite{cicchetti24}, while similarity metrics derived from pre‑trained language models have been used to quantify reconstruction fidelity \cite{Qin22,Xidong23,Jinsong24,Peng24b}. Building on these foundations, we focus on designing packet aggregation methods that ensure the transmission of core semantics in the message, through maximizing average token similarity.




In contrast to traditional packetization that treats all packets and their underlying bits with equal importance \cite{Chou07,Sundararajan11,Medard12}, this work aims to design semantics-aware packetization that groups and transmits tokens in a way that preserves the critical information in the text message even under packet losses.
To be specific, we cast packetization as a packet grouping problem and formulate it as an NP‑hard \ac{ATS} maximization problem. In our preliminary study \cite{lee2025sempa}, we proposed a greedy combinatorial optimization algorithm—termed SemPA—to find an \ac{ATS}-optimal packet group, which however incurred high computational complexity. 

To reduce the computational complexity, in this paper, we propose a novel \ac{SemPA-Look} that explicitly captures the inter-packet dependencies in sequential transmissions.
Inspired by the lookahead search algorithm \cite{snell24,Rong00,Zhang19}, our method estimates the potential contribution of future packets formed from remaining tokens, restricted to a fixed token window. This constrained prediction balances computational complexity and semantic fidelity, thereby preserving key semantics in the reconstructed message. Consequently, our proposed algorithm achieves linear complexity with respect to message length, in contrast to the exponential complexity of exhaustive search.



\subsection{Related Works}
Recent works in \ac{TC} have redefined tokens as fundamental semantic units for next-generation networks. For instance, \cite{qiao2025tcB} introduces Token-Domain Multiple Access (ToDMA) with overlapping codewords and multimodal LLM–based token assignment, while \cite{qiao2025tc} proposes a unified generative framework for semantic source compression, channel coding, and multiple access. Furthermore, \cite{Devoto25} presents an adaptive deep joint source–channel coding (DJSCC) pipeline that selects critical tokens and applies Lyapunov-based resource allocation for edge inference. These advances excel in multiple access and transceiver designs but leave packetization strategies unaddressed.

To explicitly ensure semantic preservation in text transmission, recent studies optimize end-to-end fidelity using token-level similarity metrics. For instance, R-DeepSC mitigates semantic noise with BLEU/WER evaluation \cite{Qin22}, \cite{Peng24b} define semantic spectral efficiency (S-SE) with BERT similarity, \cite{Xie21} trains Transformer-based DeepSC on sentence similarity, and \cite{Xidong23,Jinsong24} extend similarity-based designs to semi-NOMA and covert settings. While these methods improve overall message reconstruction, they do not factor individual token importance when forming packets.

In the literature of packet-level semantic communication, \cite{Yun25} proposes a synchronous multi-modal framework that interleaves words, which applies Huffman and Reed–Solomon coding and encapsulates symbols in real-time transport protocol (RTP) packets for synchronized video–text reconstruction under high-loss conditions. However, text packets are generated via uniform segmentation of the token stream, and the distinct semantic importance of individual tokens is not considered.
Similarly, \cite{Biao25} utilizes a two-dimensional wavelet transform to split packets into four subbands, allocating the low–low (LL) band to the best channel, which however targets image rate–distortion and overlooks token-level semantic grouping.

Complementing these approaches, our prior work introduced SemPA \cite{lee2025sempa}, a greedy combinatorial packet grouping algorithm that maximizes the ATS but incurs high computational complexity. We subsequently proposed SemPA-GBeam \cite{lee2025sempagbeam}, which leverages a genetic beam search to further improve ATS yet still exhibits exponential complexity with respect to the number of packets. Advancing this line of work, in this paper, we introduce a novel \ac{RSS} for quantifying each packet’s semantic importance and develop \ac{SemPA-Look} that predicts future token groupings to maximize overall semantic fidelity under erasure channels.

The proposed \ac{SemPA-Look} builds on the lookahead search algorithm \cite{snell24,Rong00,Zhang19}, which has recently been applied to various LLM-based tasks. For instance, \cite{snell24} shows that allocating more inference-time resources by looking ahead future responses improves LLM output, particularly on intermediate to difficult prompts. This result demonstrates that step-by-step look-ahead can outperform single-shot inference. 
\cite{Rong00} also shows that incorporating future observations via delayed sampling enhances sequential Monte-Carlo methods. In \cite{Zhang19}, a two-loop lookahead optimizer improves learning stability in deep neural networks. 

Compared to these lookahead applications, rather $N$ (theookahead searches, \ac{SemPA-Look} maximizes the average \ac{RSS} of packets chosen at each level. It performs a level‐wise lookahead search on a fixed token set: at each level, it enumerates candidate packets, computes average \ac{RSS} for lookaheaded samples drawn from the remaining tokens, then selects and removes the packet with the highest average RSS.

\subsection{Contributions and Paper Organization}
The main contributions of this paper are as follows:


\begin{itemize}
    \item We introduce a novel \ac{RSS} that quantifies the semantic importance of tokens within a packet by capturing the impact of token loss on the overall semantic fidelity of the transmitted message. To be specific, the \ac{RSS} is defined in a way that compares the similarity between the full set of tokens in the original message and the set obtained when a particular packet is lost or erased. This metric is computed using advanced similarity measures (e.g., cosine similarity on embeddings generated by pre-trained language models), which effectively reflect the importance of tokens for preserving the intended meaning. 

    \item Building upon the \ac{RSS}, we develop a \ac{SemPA-Look} that is capable of not only optimizing the current packet formation but also anticipating the contribution of future packets. The algorithm operates in a hierarchical lookahead manner, where at each level a candidate packet is evaluated by combining its immediate \ac{RSS} with a predicted score of potential future packets that can be constructed from the remaining tokens. In doing so, the algorithm mitigates the risk of suboptimal grouping decisions early on, thereby enhancing the overall semantic fidelity of the reconstructed message under packet erasures.

    \item Through extensive simulations, we demonstrate that our method significantly improves the token similarity between the original and reconstructed messages compared to baseline methods. Our experiments reveal that the proposed \ac{SemPA-Look} not only enhances semantic preservation by effectively leveraging information about the importance of token, but also achieves a linear computational complexity with the message length. This is in stark contrast to full search methods of exponential complexity. Thus, our approach successfully balances performance and computational efficiency, making it well-suited for practical token-based semantic communication systems.
\end{itemize}

Note that this work extends SemPA‑GBeam \cite{lee2025sempagbeam} with the following novel contributions. First, unlike SemPA‑GBeam—which achieves near-optimal \ac{ATS} but still exhibits exponential complexity in $N$(the number of transmitted packets)—we introduce a \ac{RSS} that quantifies each packet’s semantic importance by measuring the drop in overall message similarity when that packet is removed. Second, building upon RSS, we develop a \ac{SemPA-Look} that anticipates the semantic contributions of future packets, thereby mitigating early suboptimal grouping decisions. As a result, the proposed method achieves computational complexity that scales linearly with $N$.

The remainder of the paper is organized as follows. Section~\ref{sec:system_model} introduces the system model and channel assumptions for token-based communications. Section~\ref{sec:problem_formulation} formulates the average token similarity maximization problem. Section~\ref{sec:SemPA-look} presents our proposed \ac{SemPA-Look} using \ac{RSS}.
Simulation results are discussed in Section~\ref{sec:simulation_results}, and Section~\ref{sec:conclusion} concludes the paper with insights into future research directions.
The key notations used in this paper are listed in Table~\ref{tab:notations}. 

\begin{table}[t!]
\centering
\caption{List of Notations}
\label{tab:notations}
\begin{tabular}{l p{7cm}}
\toprule
\textbf{Notation} & \textbf{Definition} \\
\midrule
\(\mathcal{W}\) & The full set of tokens including the \(K\) token that makes up the text message \\

\(K\) & Total number of tokens in the text message \\

\(w_i\) & The \(i\)th token in \(\mathcal{W}\) \\

\(M\) & Number of tokens per packet \\

\(\mathcal{C}\) & A packet of tokens consisting of \(M\) tokens \\

\(\mathcal{G}\) & A packet group, satisfying \(\bigcup_{\mathcal{C}\in\mathcal{G}} \mathcal{C} = \mathcal{W}\) \\

\(N\) & Number of packets in the packet group \\

\(\phi(x, \cW)\) & Cosine similarity between the encoded text \(x\) and the original text \cW \\

\(p\) & Packet erasure probability \\

\(\psi(\mathcal{C})\) & Residual semantic score of packet \(\mathcal{C}\) reflecting its semantic importance \\

\(\mathcal{L}^{(\ell)}\) & Leftover token set available at lookahead level \(\ell\) \\
\bottomrule
\end{tabular}
\end{table}

\begin{figure*}[t]
\centering
\includegraphics[width=\textwidth]{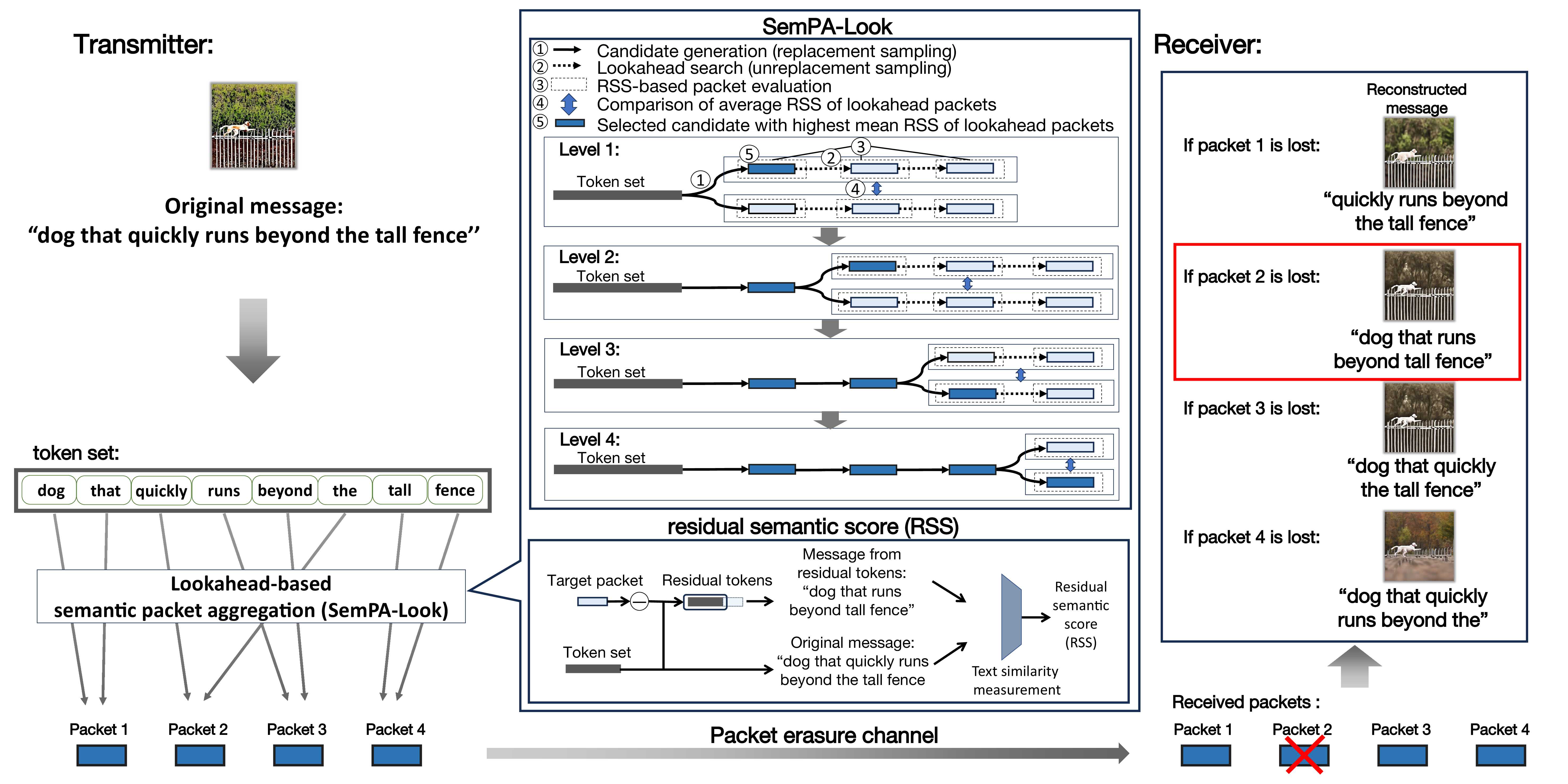}
\vspace{-1em}
\caption{The structure of \ac{SemPA-Look} for \ac{TC}. Tokens within a message are packetized, and the tokens in the received packets are reconstructed into message.}\label{fig:1}
\end{figure*}

\section{System Model}\label{sec:system_model}
\subsection{TC Scenario}
As shown in Fig.~\ref{fig:1}, for semantic communication, we consider a scenario where a text message of \(K\) tokens \([w_1, w_2, \cdots, w_K]\) is encoded and transmitted from a \ac{TX} to a \ac{RX} over an unreliable packet-level channel, where we denote \(\cW\) as the set of \(K\) tokens.
In practice, these tokens can be obtained via \emph{word-based tokenization}, where each token is simply a word split by space, or via \emph{subword-based tokenization} (e.g., byte pair encoding (BPE) \cite{Rico16}, wordpiece \cite{BERT}) that segments words into smaller units. 
In this scenario, due to a high compression ratio, the reconstruction at the \ac{RX} becomes sensitive to channel errors, i.e., if even a few tokens are lost, the \ac{RX} may not be able to reconstruct a text message that conveys the same meaning as the original.

The key distinction of our system compared to conventional packet transmission system lies in its primary objective: to maximize the semantic content of the received packets. Traditional packet transmission systems typically assume that all packets are of equal importance and focus solely on ensuring packet recovery \cite{Sundararajan11}. In contrast, since our goal is to accurately restore the semantics of the original text message, we explicitly account for the varying token importance.

Thus, to minimize the possibility that important semantics in the message is lost due to packet loss, we divide a text message of tokens into multiple packets in a way that ensures important semantics are preserved even when some packets are lost, as illustrated in Fig.~\ref{fig:1}. 
Each packet consists of $M$ tokens.
We denote the packets as \(\cC_1, \cC_2, \cdots, \cC_{N}\), where $N = \frac{K}{M}$. Here, the total number of tokens is $K$ in the sentence. We also denote \(\cG = \{\cC_1, \cC_2, \cdots, \cC_{N}\}\) as the packet group which encapsulates the whole $N$ packets needed to send the entire meaning of the sentence.
In Section~\ref{sec:problem_formulation}, we discuss how to form these packets in an optimized manner for given channel characteristics.

We consider a packet-level channel model. In practice, network-layer channels are frequently abstracted as packet-level erasure channels, reflecting how higher-layer protocols typically treat entire packets as either delivered intact or lost due to errors at lower layers \cite{Sundararajan11, Yun25}. 

We assume that each packet erasure event is independent of others.
Under the adopted model, a transmitted packet $\cC$ either arrives correctly at the \ac{RX} with probability $1-p$ or is replaced by a randomly chosen packet $\tilde{\cC}$ from the pre-defined dictionary $\cD$ to \ac{TX} and \ac{RX}, modeling the complete semantic loss of the original $M$ tokens in that packet. In addition, we assume that the original ordering of tokens within each packet \(\cC\) is fully recoverable at the \ac{RX} through the packet header information.
Denote by \({\hat{\cC}}\) the received packet at \ac{RX}, which is given as:
\begin{align}\label{Z}
{\hat{\cC}} = \begin{cases}
{\cC}, &{\text {with probability }} 1 - p, 
\\ \emptyset,  &{\text {with probability }} p,
\end{cases}
\end{align}
Here, empty set denotes that no packet is received due to the packet loss.


\subsection{Average Token Similarity (ATS)}\label{sec:ATS_problem}
The objective is to form \(\cC \in \cG\) and sequentially transmit each element as an individual packet, ensuring that the received packets encapsulate the essential semantics of the message in the presence of packet loss.
To accomplish this, we introduce a metric that quantifies the semantic similarity between texts by leveraging pre-trained large foundation models\cite{CLIP}. 
Moreover, the cosine similarity score is extensively employed as a standard measure for evaluating text similarity \cite{Guo22}, \cite{Qin22}:
\begin{align}\label{sim_f}
    \phi(x, y) = \frac{g(x)^{\mathrm{T}} g(y)}{||g(x)|| \cdot ||g(y)||},
\end{align}
where \(\phi(\cdot, \cdot)\) is the normalized token similarity function which needs two arguments as input. 

The function $g(\cdot)$ symbolizes the text encoding process of a pre-trained model that transforms inputs into feature vectors, which is frequently realized by a large foundation model or embedding encoder \cite{Xidong23, Jinsong24}. Such a step requires forward passes that typically exceed the computational cost of simpler numeric routines such as matrix multiplications or inversions. We therefore measure computational complexity in terms of \emph{the number of text encoding steps} needed, as reducing this cost is essential for scalability.

Since we need to quantify the similarity between the original message and the message decoded from any subset of received packets \(\cC \in \cG\) after packet loss, one of the argument in \(\phi(\cdot, \cdot)\) becomes \({\cW}\). 
Using \eqref{sim_f}, we can quantify the similarity between words in \({\cG}\) and \(\cW\). However, in an unreliable channel, the success of the whole packet transmission is not guaranteed. Therefore, we need another metric to evaluate the average amount of semantic information over all possible combinations of received packets and design the \ac{PA} based on this metric.

In this context, we introduce the \ac{ATS}, defined as the average similarity between the original transmitted message and each reconstructed message obtained from every possible combination of received packets.
To further explain the \ac{ATS}, consider an example where the original message is \( {\cal W} = \{\text{a}, \text{small}, \text{motor}, \text{bike}\} \), represented as \( {\cal W} = \{w_1, w_2, w_3, w_4\} \). Suppose we form two packets: \( {\cC}_1 = \{w_1, w_3\} \) and \( {\cC}_2 = \{w_2, w_4\} \).

If \({\cC}_1\) and \({\cC}_2\) are transmitted successfully, the token similarity of packets \( {\cal G} = \{{\cC}_1, {\cC}_2\} \) to the original message can be represented as follows:
\begin{align}
\phi(\cF({\cal G}), \cW) = \phi(\{w_1, w_2, w_3, w_4\}, \cW),
\end{align}
where \(\cF(\cG) =  \cup_{\cC \in \cG}\cC\). If we assume that the order of the words is given, \( \phi({\cF(\cal G)})\) is the token similarity between \(\text{``a small motor bike"}\) and \(\text{``a small motor bike"}\), which is 1.
If \ac{TX} only sends \( {\cC}_1 \) successfully, the token similarity at \ac{RX} becomes:
\begin{align}
\phi(\cF({\cC}_1), \cW) = \phi(\{w_1, w_3\}, \cW).
\end{align}
In this case, \(\phi(\cF({\cC}_1))\) is the token similarity between \(\text{``a motor"}\) and \(\text{``a small motor bike"}\).
Given the erasure probability \( p \) for each packet, the \ac{ATS} when \({\cC}_1\) and \({\cC}_2\) are sent through erasure channel is represented as:

\begin{equation}\label{eq:ATS}
\begin{split}
{\mathbb E}\!\left[\phi(\cF(\cG),\!\cW)\right]\!&= \!(1\!-\!p)^2\!\!\cdot\!\phi(\cF({\cal G}),\!\cW)\!+\!(1\!-\!p) p\!\cdot\!\phi(\cF({\cC}_1),\!\cW)\! \\ 
&+\!(1\!-\!p) p\!\cdot\!\phi(\cF({\cC}_2),\!\cW).
\end{split} 
\end{equation}

In more general terms, if \({\cG} = \{{\cC}_{1},\cdots, {\cC}_{N}\} \), the \ac{ATS} is given as:

\begin{align}
\hspace{-0.5pc}{\mathbb E}\left[\phi(\cF(\cG), \cW)\right] = \sum_{\cH \subseteq \cG}\;(1 - p)^{|\cH|}p^{N - |\cH|}\cdot \phi\bigl(\mathcal{F}(\mathcal{H}), \cW\bigr),
\end{align}
where \(\cH\) represents any possible subset of \({\cG}\).

\section{Problem Formulation}\label{sec:problem_formulation}
\subsection{ATS Maximization Problem}
In this section, we formulate an \ac{ATS} maximization problem in the presence of packet erasures. We assign each token in \(\cW\) into one of the packets \(\cC \in \cG\). Since packets are lost randomly, we denote by \( \cH \) the subset of \( \cG \) representing received packets in each loss scenario and aim to maximize the average semantic similarity across all possible \( \cH\) configurations.
Each \(\cC \in \cG\) is transmitted once as a single packet.
The problem for maximizing the \ac{ATS} can be formulated as:
\refstepcounter{equation}\label{P0}
\begin{flalign*}
\textbf{P1: }&\max \limits_{\cG}{\mathbb E}\left[\phi(\cF(\cG),\cW)\right]\tag{\theequation a}\label{P0a} &&
\\& {~{\text {s.t.}}}\hspace{0.3pc} \cF({\cG})\!=\!{\cW}\;({\text{complete transmission constraint}}),\hspace{-0.5pc}\tag{\theequation b}\label{P0b}&&
\\& \hphantom{~{\textrm {s.t.}}}\hspace{0.3pc}|\cC| = M\hspace{0.3pc},\forall \cC \in \cG \;({\text{packet length constraint}}), \tag{\theequation c}\label{P0c}&&
\end{flalign*}
where the objective in \eqref{P0a} is defined according to the \ac{ATS} metric introduced in \eqref{eq:ATS}. The constraint in \eqref{P0b} ensures that the entire set of tokens \(\cW\) must be delivered, reflecting the practical transmission goal of the \ac{TX} to fully convey the semantic content. The packet length constraint in \eqref{P0c} reflects the fact that most communication protocols in practice commonly utilize uniform-sized packets for protocol design and network implementation \cite{Stevens12}.




Finding the optimal solution of \(\cG\) in {\textbf{P1}} is computationally challenging due to dependencies among packets. Such dependencies naturally arise because packets are transmitted sequentially. For the message reconstruction, the message is recovered by aggregating information from all received packets. Consequently, the optimal token composition to be included in the current packet varies depending on the transmission success of the previously sent packets. 
Moreover, similarity of multiple tokens is not directly derived from individual token similarities due to complex contextual dependencies between tokens, making \textbf{P1} an NP‑hard combinatorial problem.
In particular, directly solving \textbf{P1} by optimizing $\cG$ can be distilled into two key issues:

\begin{itemize}
  \item \textbf{Exponential Number of Feasible Packet Groups.}
    Since our objective function relies on a non-linear metric $\phi(\cdot)$ derived from an encoder $g(\cdot)$ (e.g., a large foundation model), there is no closed-form or straightforward approach to optimize \eqref{P0a} without explicitly comparing different packet group $\cG$. In other words, gradient–based methods or polynomial-time combinatorial algorithms cannot be applied.
    Consequently, if one were to attempt a full search to find the optimal $\cG$, every possible partitions of $K$ tokens into disjoint packets of size $M$ would need to be enumerated and evaluated. The total number of such packet groups is $\frac{K!}{(M!)^{N}\cdot \left(N\right)!}$,
    which grows rapidly with $K$ and $N$. This combinatorial complexity growth makes a na\"ive full search infeasible in practice, requiring an exponential number of processing the non-linear encoder $g(\cdot)$ to compare all possible $\cG$. Hence, we need to resort to approximation or low-complexity heuristic methods for efficient optimization.
  \item \textbf{Cost of Text Encoding Steps for Evaluating packet group.}
    Even if one enumerates a single candidate \(\cG\), calculating the \ac{ATS} by \eqref{P0a} is extremely costly, as it involves computing \(g(\cdot)\) for all subsets in \(\cG\) to evaluate \ac{ATS}.
    To be concrete, each \(\cG\) evaluation entails the number $2^{N}$ of text-encoding steps, leading to a dominant bottleneck for large $N$.
  \end{itemize}

Consequently, effective yet computationally efficient heuristic or approximation algorithms are necessary to solve {\textbf{P1}}.

\subsection{Baseline: Genetic Algorithm Based Packet Aggregation}\label{sec:GA}
Reducing the computational complexity of \textbf{P1} is non-trivial. To address its difficulty, in this subsection, we consider a \ac{GA} approach, a well-known evolutionary algorithm that iteratively refines solutions through selection, crossover, and mutation operations, converging on a near-optimal solution \cite{katoch21}. 

To be specific, each chromosome represents as a random set of $N$ packets of fixed size $M$ drawn from the overall token set $\mathcal{W}$. For each generation, we evaluate the fitness of each chromosome by computing its exact \ac{ATS}. Then crossover and mutation are performed to introduce diversity in the population:

\begin{enumerate}
\item \textbf{Crossover}
Suppose each of two parent chromosomes, $\cG_1^{(p)}$ and $\cG_2^{(p)}$, is represented by a list of packets, i.e.,
\begin{align}
  \cG_1^{(p)} 
  &\;=\; [\,\cC_1^{(1)}, \cC_2^{(1)}, \dots, \cC_{N}^{(1)}\bigr], 
  \label{eq:parent1}
  \\
  \cG_2^{(p)} 
  &\;=\; [\,\cC_1^{(2)}, \cC_2^{(2)}, \dots, \cC_{N}^{(2)}\bigr].
  \label{eq:parent2}
\end{align}
Each $\cC_j^{(\ell)}$ is a size-$M$ subset of $\{0,\dots,K-1\}$ corresponding to a packet of tokens. 
A simple crossover can be carried out as follows:
    \begin{enumerate}
        \item \emph{Random Cut}: pick an integer $c \in \{1,2,...,N-1\}$, then define
        \begin{align}
          \cG_1^{(c)} &= 
            [\,\cC_1^{(1)}, \dots, \cC_l^{(1)}\bigr]
            \,\|\, 
            [\,\cC_{l+1}^{(2)}, \dots, \cC_{N}^{(2)}\bigr], 
          \\
          \cG_2^{(c)} &= 
            [\,\cC_1^{(2)}, \dots, \cC_l^{(2)}\bigr]
            \,\|\, 
            [\,\cC_{l+1}^{(1)}, \dots, \cC_{N}^{(1)}\bigr],
        \end{align}
        where $\|$ denotes concatenation. 
        \item \emph{Repair Overlaps}: if any duplicate indices arise within a child, or if the total flattened set has size different from $K$, a repair routine is applied to re-randomize or fix conflicts. 
        \item \emph{Outcome}: two child chromosomes, each still a list of $N$ size-$M$ packets, are produced as follows:
        \begin{equation}
          \bigl(\cG_1^{(c)}, \cG_2^{(c)}\bigr).
        \end{equation}
    \end{enumerate}


\item \textbf{Mutation}
To preserve the size-$M$ constraint for each packet, a simple mutation rule is adopted:
    \begin{enumerate}
        \item \emph{Packet selection}: For a chromosome $[\cC_1, \cC_2, \dots, \cC_{N}]$, select one packet $\cC_m$ at random.
        \item \emph{Tokens swapping:}
    Choose one token $w_{\mathrm{old}}$ from $\cC_m$ and one token $w_{\mathrm{new}}$ from $\cC_n$. Formally,
\refstepcounter{equation}
\begin{align}
&\cC_m' = \bigl(\cC_m \setminus \{w_{\mathrm{old}}\}\bigr) \cup \{w_{\mathrm{new}}\},
\tag{\theequation a}\\
&\cC_n' \;= \bigl(\cC_n \setminus \{w_{\mathrm{new}}\}\bigr) \cup \{w_{\mathrm{old}}\}. \tag{\theequation b}
\end{align}

    \item \emph{Update:}
    Replace $(\cC_m, \cC_n)$ with $(\cC_m',\cC_n')$ in the chromosome. Both $\cC_m'$ and $\cC_n'$ remain of size $M$, preserving the packet-length constraint.
    \end{enumerate}
\end{enumerate}

We repeat this process for $G$ generations using a population size $P$. \ac{GA} effectively addresses the first challenge by iteratively searching only a subset of feasible packet groups, thus avoiding exhaustive comparisons of feasible $\cG$. However, the second challenge of solving $\textbf{P1}$ still persists, because evaluating the \ac{ATS} for each chromosome still requires computing similarity scores over all subsets of packets. Consequently, the total complexity remains at $G\cdot P\cdot 2^N$ text encoding steps, making \ac{GA} impractical for large $N$.

In this context, we propose to reduce the complexity of finding optimal $\cG$ by evaluating the individual packets and approximate \eqref{P0a} instead of computing it directly, by introducing the surrogate function.


\section{Proposed Approach: lookahead-based Semantic Packet Aggregation (SemPA-Look)}\label{sec:SemPA-look}

\subsection{Problem Reformulation via a Surrogate Function}\label{sec:problem_reformulation}
As discussed in Sec.~\ref{sec:ATS_problem}, the one of the challenges in maximizing \eqref{P0a} arises from computing \ac{ATS} by summing over $2^N$ subsets of $\cG$.
To avoid these excessive text encoding steps, we propose a surrogate function for each packet $\mathcal{C}$, i.e., $\psi(\cC, \cW)$, allowing us to evaluate its contribution with only \emph{one} text encoding step per packet. By assessing each packet $\cC$ individually, the total number of required text encoding steps to evaluate the packet group becomes $N$, rather than $2^N$.

Consequently, we can reformulate \textbf{P1} into \textbf{P2} that seeks to maximize the average of the surrogate functions of individual packets as follows:
\begin{flalign*}
\textbf{P2: }&\max_{\{\cC_i\}_{i=1}^{N}} \frac{1}{N} \sum_{i=1}^{N} \psi(\cC_i, \cW)\tag{\theequation a}\label{P3a} &&
\\&\hspace{0.35pc}{~{\text {s.t.}}}\quad \cF({\cG})\!=\!{\cW}\tag{\theequation b}\label{P3b}&&
\\& \hphantom{~{\textrm {s.t.}}}\hspace{0.5cm}|\cC_i| = M \quad \forall\, i=1,\ldots,N. \tag{\theequation c}\label{P3c}&&
\end{flalign*}
In \textbf{P2}, the packet group $\cG$ is decoupled and each packet $\cC_i$ is allowed to be evaluated independently.

Having reduced \textbf{P1} into \textbf{P2}, the next challenge is choosing a surrogate function that can approximate \eqref{P0a}. The choice of an effective surrogate function depends significantly on the packet decoding error rate, $p$. Intuitively, if the packet decoding error rate \(p\) is high, it becomes crucial that any single successfully received packet contains enough important semantic information to reconstruct the context of the sentence. On the other hand, when $p$ is relatively low, the focus shifts to ensuring that, no matter which single packet is lost, the information contained in the remaining packets preserves as much of the original message’s semantics as possible. In this light, we can distinguish two principal strategies based on channel conditions:

\begin{enumerate}
  \item \textbf{Top semantic score (TSS):}
    When $p$ is high, each surviving packet must individually contain enough semantic information to reconstruct a meaningful portion of the original sentence. Then, a straightforward, greedy approach can be employed, which prioritizes individual token importance:
    \begin{align}
    \psi(\mathcal{C}_i, \cW) = \phi\bigl(\mathcal{C}_i, \cW\bigr).
    \end{align}
    
    However, when packet decoding error rate $p$ is low, if $\mathcal{C}_i$ consists solely of highly important tokens, then losing $\mathcal{C}_i$ leads to a substantial drop in the \ac{ATS}. 

    \item \textbf{Residual semantic score (RSS):} Conversely, when the packet erasure probability $p$ is low, our goal shifts to maximizing the token similarity of whatever remains after any single packet loss. To achieve this, we maximize the similarity obtained by reconstructing the message from the remaining packets when any packet $\cC_i$ is lost. Rather than evaluating individual packet $\cC_i$, we assess the impact of removing $\cC_i$ by measuring the similarity between the original sentence and the reconstruction formed from all other packets as follows:
    \begin{align}
    \psi(\mathcal{C}_i, \cW) = \phi\bigl(\mathcal{W}\setminus \mathcal{C}_i, \cW\bigr).
    \end{align}
\end{enumerate}

\begin{table}[t!]
    \caption{Comparison of ATS for RSS‑based PA, TSS‑based PA, and Random PA on the example sentence ``dog that quickly runs over the tall fence.''
}
    \label{tab:performance_side}
    \centering
    \setlength{\tabcolsep}{3pt}  
    \begin{tabular}{lccc}
        \toprule
        \textbf{ } & \textbf{RSS-based PA} & \textbf{TSS-based PA} & \textbf{Random PA} \\
        \midrule
        ATS    & 0.8397  & 0.7708 & 0.8091 \\
        \bottomrule
    \end{tabular}
\end{table}
\begin{figure}[t]
\centering
\includegraphics[width=0.8\columnwidth]{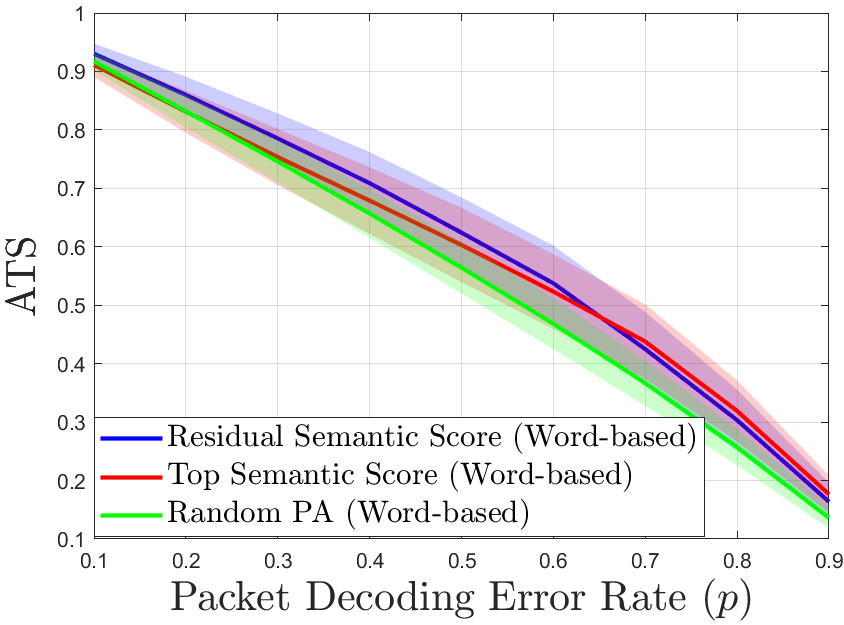}
\caption{Cosine Similarities of different \ac{PA} strategies. The shaded areas indicate the ±1 standard deviation uncertainty bands around the mean values.}\label{fig:plot_p_inv}
\end{figure}


Table~\ref{tab:performance_side} compares the ATS of three \ac{PA} for the sentence ``dog that quickly runs beyond the tall fence'' under \(p=0.25\): RSS‑based PA, TSS‑based PA, and Random PA. RSS‑based PA achieves an ATS of 0.8397, Random PA yields 0.8091, and TSS‑based PA attains only 0.7708, the lowest among the three. These results show that TSS‑based grouping underperforms even random \ac{PA}, while RSS‑based \ac{PA} most effectively maximizes ATS.

Furthermore, Fig.~\ref{fig:plot_p_inv} compares the \ac{ATS}s of three \ac{PA} under a wide range of $p$. Under the practical channel condition ($p<0.6$), RSS-based \ac{PA} consistently yields higher \ac{ATS} than TSS-based \ac{PA}. We therefore adopt \ac{RSS} as the surrogate function for evaluating individual packet \(\cC_i\).



Although the surrogate function is designed to avoid the direct optimization of $\cG$, when searching for \(\cC_i \) to be included in \(\cG\), the remaining tokens and other \(\cC_j \in \cG\) should also be considered. This is because each packet faces an independent probability of failure, and the combinations of successfully transmitted packets at the end collectively reconstruct the text message. Hence, the token similarities of the collected tokens in the subsets of other \(\cC_j \in \cG\) and \(\cC_i \) should be taken into account.

To address these challenges, we propose \ac{SemPA-Look} to find \(\cC_i \in \cG\) that maximizes the \ac{ATS}. At each level of the algorithm, we generate candidate packets as nodes, evaluate them using a proposed \ac{RSS}, and approximate the potential contribution from the future packets that could be formed using those leftover tokens. 



\subsection{Impact of packet length on packet loss and semantic loss}\label{sec:trade_off}
To understand how packet length $M$ affects token similarity, we analyze the tradeoff between packet loss and semantic loss through approximations for different range of $p$.
When $p$ is high, i.e., $p \approx 1$, \ac{ATS} primarily depends on scenarios where only a small number of packets are successfully received. Using the first-order taylor expansion, \ac{ATS} is approximated as:
\begin{align}\label{eq:high_p_taylor}
&\sum_{\mathcal{H}\,\subseteq\,\mathcal{G}}
(1-p)^{\lvert\mathcal{H}\rvert}\,
p^{\,N - \lvert\mathcal{H}\rvert}\,
\phi\Bigl(\mathcal{F}\bigl(\mathcal{H}\bigr),\;\mathcal{W}\Bigr)
\\
&=\!(1\!-\!p)^{\frac{K}{M}}\phi(\emptyset,\!\cW)\! +\!(1-p)p\!^{\frac{K}{M}-1}\! \sum_{\mathcal{C} \in \mathcal{G}} \phi(\mathcal{C}, \cW)\!+\!{\mathcal{O}}((1\!-\!p)^2)
\\&\approx(1\!-\!p)\!\left(\!1\!-\!\left(\frac{K}{M}\!-\!1\right)(1-p)\!\right)\!\sum_{\mathcal{C} \in \mathcal{G}} \phi(\mathcal{C}, \cW)\!+\!{\mathcal{O}}((1-p)^2)
\\&\approx\!\left(\!(1\!-\!p)\!+\!\left(\frac{K}{M}\!-\!1\right)(1-p)^2\!\right)\!\sum_{\mathcal{C} \in \mathcal{G}} \phi(\mathcal{C}, \cW)\!+\!{\mathcal{O}}((1-p)^2)
\\&\approx (1\!-\!p)\sum_{\mathcal{C} \in \mathcal{G}} \phi(\mathcal{C}, \cW),
\end{align}
From \eqref{eq:high_p_taylor}, \ac{ATS} dominantly depends on the semantic content of individual packets.

Conversely, for low $p$, the expected similarity depends on $M$ in more complicate manner as follows:
\begin{align}\label{eq:low_p_taylor}
\hspace{-0.2pc}&\sum_{\mathcal{H}\,\subseteq\,\mathcal{G}}
(1-p)^{\lvert\mathcal{H}\rvert}\,
p^{\,N - \lvert\mathcal{H}\rvert}\,
\phi\Bigl(\mathcal{F}\bigl(\mathcal{H}\bigr),\;\mathcal{W}\Bigr)
\\
&=(1-p)^{\frac{K}{M}}+(1-p)^{\frac{K}{M}-1}p \sum_{\mathcal{C} \in \mathcal{G}} \phi(\mathcal{W}\setminus\mathcal{C}, \cW)+ {\mathcal{O}}(p^2) \hspace{-0.5pc}
\\&\approx (1-\frac{K}{M}p)\!+\!\left(p -(\frac{K}{M} - 1)p^2\!\right) \!\!\sum_{\mathcal{C} \in \mathcal{G}} \phi(\mathcal{W}\setminus\mathcal{C}, \cW) \!+ \!{\mathcal{O}}(p^2)
\\&\approx(1-\frac{K}{M}p)+ p \sum_{\mathcal{C} \in \mathcal{G}} \phi(\mathcal{W}\setminus\mathcal{C},\cW) + {\mathcal{O}}(p^2)
\\&\approx 1-p \bigg(\underbrace{\frac{K}{M}}_{\text{packet loss}}\underbrace{- \sum_{\mathcal{C}\in \mathcal{G}}
\phi\bigl(\mathcal{W}\setminus \mathcal{C},\mathcal{W}\bigr)}_{\text{semantic loss}} \bigg)
\end{align}

In \eqref{eq:high_p_taylor}, \ac{ATS} is approximated by the sum of token similarities of individual packets. Conversely, \eqref{eq:low_p_taylor} comprises a packet loss term and a semantic loss term. As $M$ increases, the packet loss term $\frac{K}{M}$ decreases, whereas the semantic loss increases. Thus, packet length $M$ induces an inherent trade‑off between packet loss and semantic loss. We will analyze this trade‑off in Section~\ref{sec:simulation_results}.
\subsection{Lookahead-based Semantic PA (SemPA-Look)}\label{sec:lookahead}
By reformulating \textbf{P1} into \textbf{P2}, the complex task of directly optimizing the packet group $\cG$ is decomposed into selecting individual packets ${\cC}_i$. This decomposition facilitates the use of an efficient lookahead algorithm. To be specific, sequential selection of individual packets ${\cC}_i$ depends on previously selected ones. In this sequential decision scenario, the lookahead algorithm \cite{snell24} fits in naturally by predicting upcoming packets, thereby preventing early decisions that may seem optimal at the time, but ultimately lead to poorer overall performance.


Our proposed \ac{SemPA-Look} predicts future packets to be aggregated, thereby avoiding suboptimal early choices that might aggregate important tokens together in a single packet.
To be specific, the key insight for using a \ac{SemPA-Look} is that one packet is decided for every level, while considering the possible leftover tokens for future packets. Rather than selecting ${\cC}_i$ for $\cG$ independently, each candidate tuple is evaluated by combining its own metric with packet sampling from leftover token set, thus capturing how the leftover tokens might form additional packets.

At each level of the algorithm, $\ell = 1, 2, \dots$, we aim to choose one packet $\mathcal{C}_{\ell}$. We define $\mathcal{L}^{(\ell)}$ as the \emph{leftover} tokens available at level $\ell$, i.e.,
\begin{equation}
  \mathcal{L}^{(1)} \;=\; \mathcal{W}, 
  \quad
  \mathcal{L}^{(\ell)} \;=\; \mathcal{W}\,\setminus\,\bigcup_{u=1}^{\ell-1}\mathcal{C}_{u} 
  \quad (\ell \ge 2).
\end{equation}
Thus, at level $\ell$, we focus on forming a new packet $\mathcal{C}_{\ell} \subseteq \mathcal{L}^{(\ell)}$ with $\vert\mathcal{C}_{\ell}\vert = M$.

We define a \emph{lookahead packet sampling} procedure as follows:
Given a candidate packet $\tilde{\cC}^{(\ell)}_{i} \subseteq \mathcal{L}^{(\ell)}$ at level $\ell$, let
  \begin{equation}
    \widetilde{\mathcal{L}}_i \;=\; \mathcal{L}^{(\ell)} \,\setminus\, \tilde{\cC}^{(\ell)}_{i}.
  \end{equation}
  Then we randomly sample $k$ packets $\bigl(\hat{\mathcal{C}}^{(i)}_{1}, \dots, \hat{\mathcal{C}}^{(i)}_{k}\bigr)$ from $\widetilde{\mathcal{L}}_i$ such that each $\hat{\mathcal{C}}^{(i)}_{1},..., \hat{\mathcal{C}}^{(i)}_{k}$ is disjoint from the others, i.e., $\hat{\mathcal{C}}^{(i)}_j \cap \hat{\mathcal{C}}^{(i)}_{j'} = \emptyset$ for all $j \neq j'$, and each $\vert \hat{\mathcal{C}}^{(i)}_j \vert = M$. We denote this operation by
  \begin{equation}
    \{ \hat{\mathcal{C}}^{(i)}_{1}, \dots, \hat{\mathcal{C}}^{(i)}_{k} \} \;\;=\;\mathrm{RandomDisjoint}(\widetilde{\mathcal{L}}_i, M, k).
  \end{equation}
  
For lookahead packet sampling in the \ac{SemPA-Look}, instead of applying replacement strategy, the sampling with unreplacement strategy is adopted to sample future packets from the leftover tokens. Such an approach is often used in backtracking algorithm to ensure that the sampled variables can satisfy the constraint of the problem. In \eqref{P0}, the final set $\cG$ must form a valid partition of the overall token set $\cW$, meaning that future packets to be selected are required to be disjoint packets. 

\begin{figure}
\centering
\includegraphics[width=0.8\columnwidth]{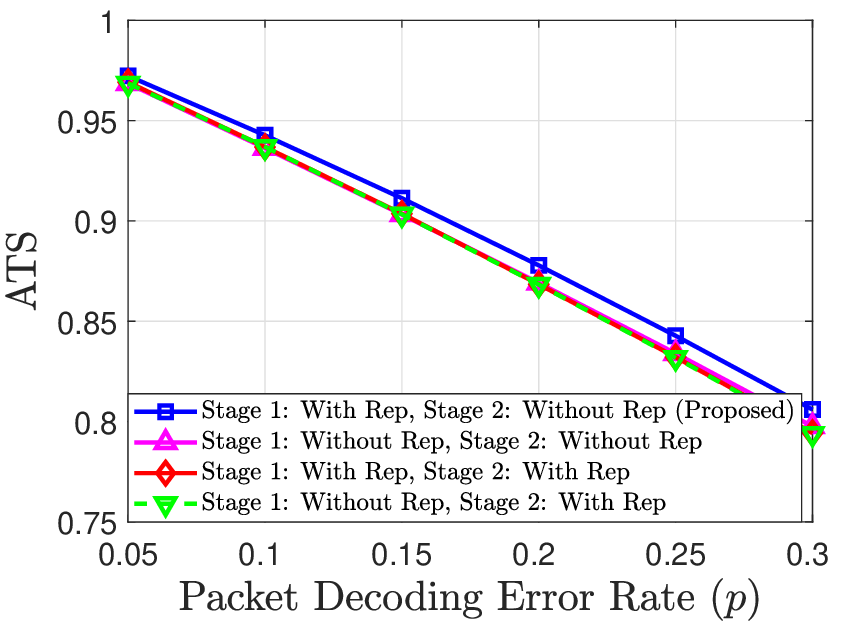}
\caption{\ac{ATS} of \ac{SemPA-Look} with different sampling strategies for Stage 1: candidate generation and Stage 2: lookahead packet sampling. $K=12, M=2, P = 10, k = 4$.}\label{fig:plot_rep}
\end{figure}

On the other hand, we adopt the sampling with replacement strategy for candidate tuples generation to explore a wide variety of potential packets. Hence, during this phase, some candidate packets may share duplicate tokens. The simulation results in Fig.~\ref{fig:plot_rep} shows that the proposed sampling method for lookahead packet sampling and candidate tuples generation is effective.

Subsequently, the average of \ac{RSS}s $\psi(\tilde{\mathcal{C}}^{(\ell)}_{i}, \cW),$$ \psi(\hat{\mathcal{C}}^{(i)}_1, \cW),$ $ \dots,$ $ \psi(\hat{\mathcal{C}}^{(i)}_k, \cW)$ is computed as follows:
  \begin{equation}\label{eq:k-step-sum}
    \hspace{-0.5pc}\Psi(\tilde{\mathcal{C}}^{(\ell)}_{i}, \cW) \!=\! \frac{1}{k+1} \left(\psi(\tilde{\mathcal{C}}^{(\ell)}_{i}, \cW) \;+\; \sum_{j=1}^k \psi(\hat{\mathcal{C}}^{(i)}_j, \cW)\right).
  \end{equation}
Essentially, $\Psi(\tilde{\mathcal{C}}^{(\ell)}_{i}, \cW)$ approximates the potential contribution of future packets when $\tilde{\mathcal{C}}^{(\ell)}_{i}$ is selected, by evaluating $k$ randomly sampled disjoint packets from the leftover tokens.

\begin{figure}[t]
\centering
\includegraphics[width=.9\columnwidth]{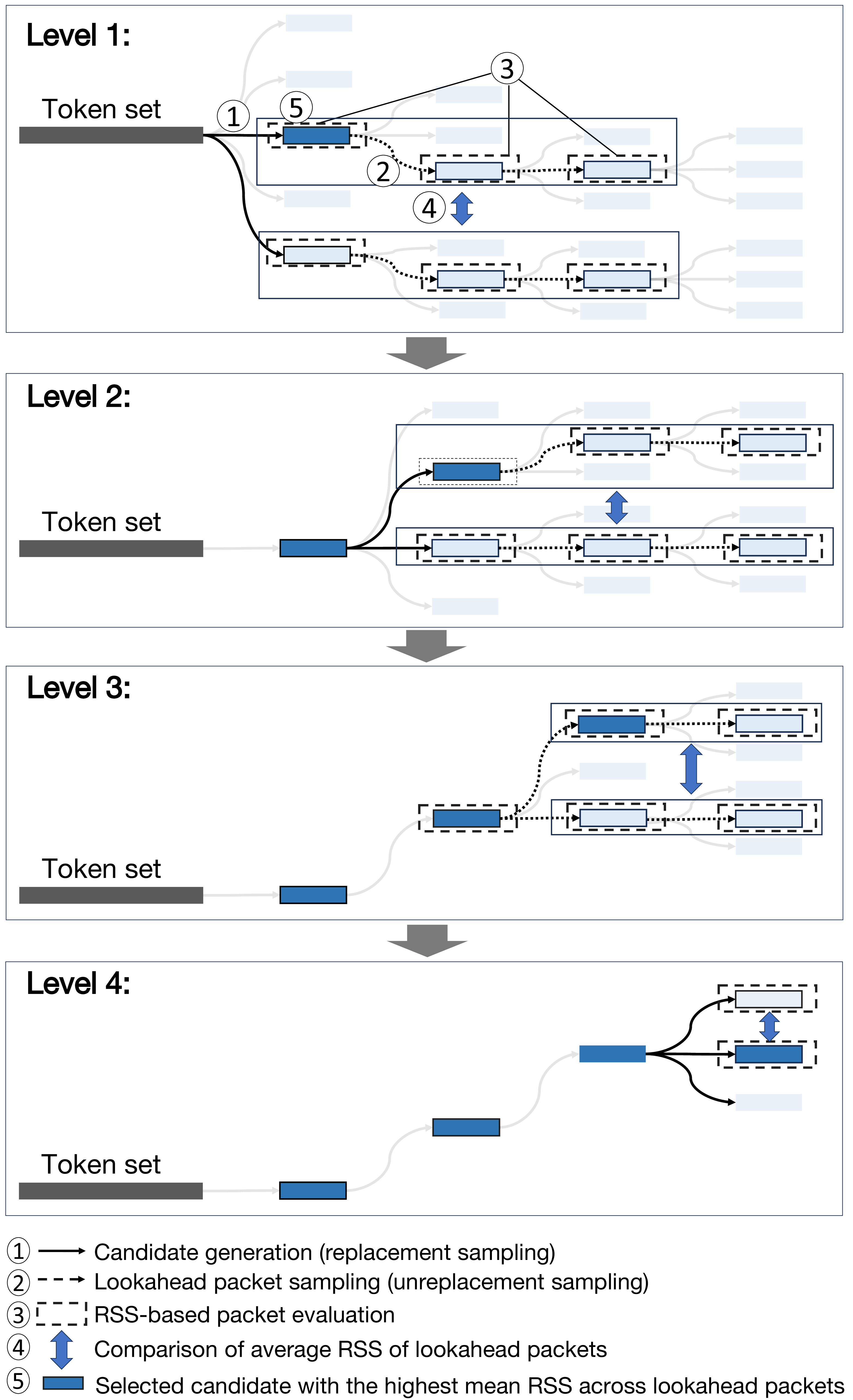}
\caption{An illustration of the \ac{SemPA-Look} algorithm.}\label{fig:tree_search_alg}
\end{figure}

With these preliminaries, the five stages executed at each level \(\ell\) (\(\ell \ge 1\)) are detailed in Algorithm~\ref{alg:packet_agg} and Fig.~\ref{fig:tree_search_alg}. In summary, at each level the algorithm selects a single packet \(\mathcal{C}^{(\ell)}_{i^*}\) by combining its immediate RSS with RSSs from its \(k\)-step lookahead packets. Iterating this procedure yields the final packet group \(\cG = \{\mathcal{C}_1, \mathcal{C}_2, \dots, \mathcal{C}_N\}\) designed to maximize \ac{ATS}.

\begin{algorithm}[t]
\caption{\small \ac{SemPA-Look}}\label{alg:packet_agg}\small 
\begin{algorithmic}[1]
\Require Token set $\cW$, packet size $M$, lookahead depth $k$, candidates $P$, packets $N$
\State $\ell \leftarrow 1$, $\mathcal{L}^{(1)} \leftarrow \cW$
\While{$\ell \le N$ and $\mathcal{L}^{(\ell)} \neq \emptyset$}
    \Statex \;\; \textcircled{1} \textbf{Candidate Generation}
    \State Sample $P$ packets $\{\cC_i^{(\ell)}\}_{i=1}^P$ of size $M$ from $\mathcal{L}^{(\ell)}$
    \Statex \;\; \textcircled{2} \textbf{Lookahead Packet Sampling}
    \For{$i = 1$ to $P$}
        \State $\widetilde{\mathcal{L}}_i \leftarrow \mathcal{L}^{(\ell)} \setminus \cC_i^{(\ell)}$
        \State Sample disjoint packets $\{\hat\cC_j^{(i)}\}_{j=1}^k$ from $\widetilde{\mathcal{L}}_i$
    \EndFor
    \Statex \;\; \textcircled{3} \textbf{RSS-Based Evaluation}
    \For{$i = 1$ to $P$}
        \State Compute $\psi(\cC_i^{(\ell)},\cW)$ and $\psi(\hat\cC_j^{(i)},\cW)$ for $j=1,\dots,k$
    \EndFor
    \Statex \;\; \textcircled{4} \textbf{Average RSS Computation}
    \For{$i = 1$ to $P$}
        \State $\Psi_i \leftarrow \tfrac{1}{k+1}\bigl(\psi(\cC_i^{(\ell)},\cW) + \sum_{j=1}^k \psi(\hat\cC_j^{(i)},\cW)\bigr)$
    \EndFor
    \Statex \;\; \textcircled{5} \textbf{Candidate Selection}
    \State $i^* \leftarrow \arg\max_{i}\,\Psi_i$
    \State Select $\cC_\ell \leftarrow \cC_{i^*}^{(\ell)}$
    \State $\mathcal{L}^{(\ell+1)} \leftarrow \mathcal{L}^{(\ell)} \setminus \cC_{i^*}^{(\ell)}$
    \State $\ell \leftarrow \ell + 1$
\EndWhile
\State \Return Packet grouping $\{\cC_1,\dots,\cC_\ell\}$
\end{algorithmic}
\end{algorithm}

\section{Simulation Results}\label{sec:simulation_results}
We evaluate the proposed \ac{SemPA-Look} by comparing them with baseline methods. We assume that the tokens of the text message are included in the pre-defined dictionary \( \cD = \{v_1, v_2, \cdots, v_{|\cD|}\}\) in both \ac{TX} and \ac{RX}. The tokens in each packet are encoded to indices in the dictionary, which needs \(\lceil \log_2 |\cD| \rceil\) bits for each token. 
The token similarity is measured by computing cosine of the angles between two text embeddings which are derived from the pre-trained \ac{CLIP} model. 

To thoroughly evaluate the performance of the proposed \ac{SemPA-Look} and examine the trade‑offs arising from the settings (e.g., packet size), we conduct experiments on messages from two distinct datasets. First, we examine the performance using MS-COCO dataset \cite{MSCOCO14}. Second, we analyze the impact of the \ac{PA} using Wiki-how dataset \cite{koupae18}. 
Finally, we turn our attention to different task, \ac{AIGC} experiments in which reconstructed sentences are used to generate new images; through these perceptual evaluations, we analyze how effectively the semantic content of the captions is preserved under packet loss and how this affects \ac{AIGC} task such as text-to-image synthesis.

We describe four baseline methods to solve {\textbf{P1}}, in addition to \ac{GA}. 



\begin{itemize}
    \item \textbf{Random \ac{PA}:} $\cG$ is generated as a random partition of $\cW$ that satisfies \eqref{P0b} and \eqref{P0c}. 
    This method incurs no computational cost, although it is expected to yield a relatively low token similarity.
    
    \item \textbf{Full Search:} This method exhaustively enumerates all feasible packet sets $\cG$ that satisfy the constraints in \eqref{P0b} and \eqref{P0c}, then selects the optimal packet set $\cG^{\star}$ by maximizing the \ac{ATS}:
    \[
    \cG^{\star} = \argmax_{\cG}{\mathbb E}\left[\phi(\cF(\cG),\cW)\right].
    \]
    Although this approach guarantees the optimal solution for \textbf{P1},  it suffers from two critical sources of computational complexity: the number of feasible $\cG$ grows according to $\frac{K!}{(M!)^{N} \cdot (N)!}$, and evaluating the \ac{ATS} for each such $\cG$ requires $2^N$ text encoding steps.
    
    \item \textbf{SemPA-GBeam \cite{lee2025sempagbeam}:} 
    \ac{SemPA-GBeam} is a semantic‐aware \ac{PA} that integrates beam search and GA: at each iteration it retains the top $B$ packet group candidates, generates $P/B$ mutated childs from each (for a total of $B\times\frac{P}{B}=P$ groups), and then selects the best $B$ packet groups for the next iteration.
    It achieves performance close to Full Search with same computational complexity as \ac{GA}.
    
    \item \textbf{Without Packetization:} In this method, all tokens \(w \in \cW\) are aggregated into a single packet, and the entire sentence is transmitted as one full packet. This method increases the probability of complete transmission of message, however, it is highly vulnerable to the losing whole tokens.
\end{itemize}


We present a detailed performance analysis of our approach on the MS-COCO dataset, focusing on how different tokenization methods (word-based versus subword-based), diverse packet decoding error rates $p$, and various packet sizes $M$ affect the token similarities of the reconstructed sentence from received packets and the original sentence.

\begin{figure}
     \centering
     \subfloat[Word-based tokenization is applied.]{%
         \includegraphics[width=0.48\columnwidth]{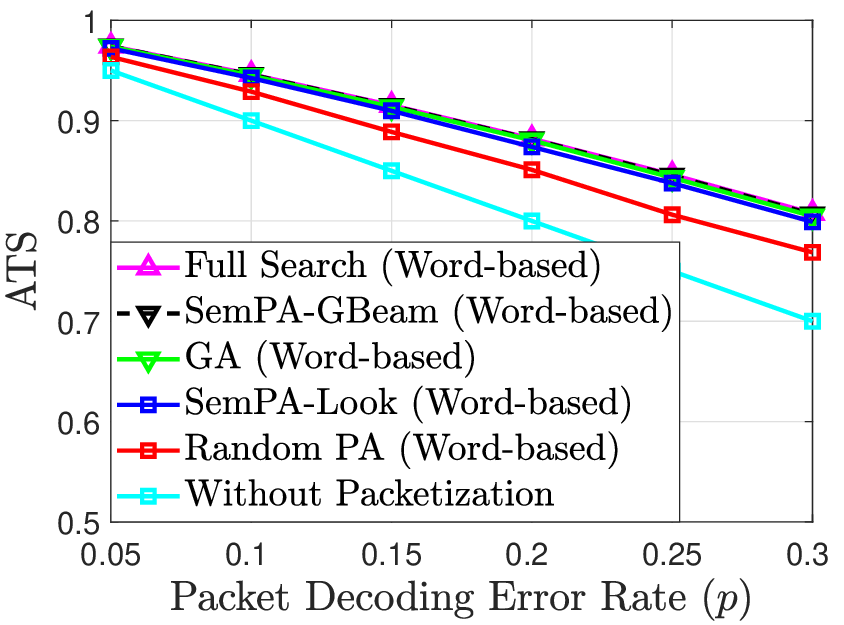}
         \label{fig:plot_p_word}
     }
     \hfill
     \subfloat[Subword-based tokenization is applied.]{%
         \includegraphics[width=0.48\columnwidth]{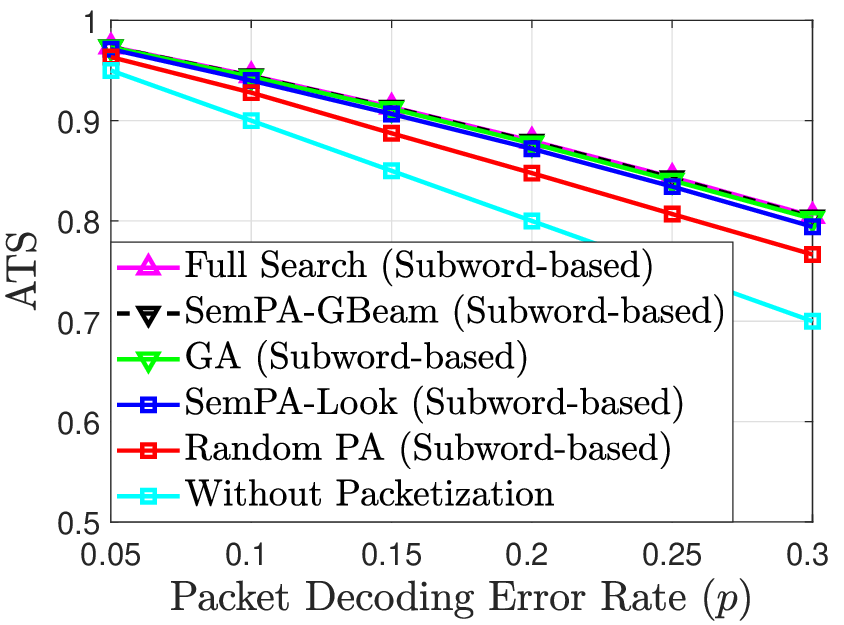}
         \label{fig:plot_p_token}
     }
     \caption{\ac{ATS} of different PA methods over different \(p\). \ac{SemPA-Look} consistently attains near-optimal ATS over a full range of $p$ ($K = 8, M = 4, P = 10,  k = 4).$} 
     \label{fig:plot_p}
\end{figure}

\vspace{1em}
\noindent \textbf{\ac{ATS} Performance of \ac{PA} Methods on the MS-COCO Dataset}.\quad
To identify the robustness of the proposed \ac{SemPA-Look} over a wide range of channel condition, in Fig.~\ref{fig:plot_p}, we analyze the impact of different \ac{PA} by comparing \ac{ATS} for various $p$.
In Fig.~\ref{fig:plot_p_word}, \ac{SemPA-Look} attains nearly the same score as the optimal full search and the near-optimal SemPA-GBeam, but slightly lower \ac{ATS}. 
Considering that other methods have high computational complexity which exponentially increases with $K$ or $N$ as shown in Tab.~\ref{tab:complexity}, \ac{SemPA-Look} is shown to be computationally efficient, where the computational complexity linearly increases with $K$, yet capable of reaching high \ac{ATS}. 
Fig.~\ref{fig:plot_p} clearly demonstrates that \ac{SemPA-Look} achieves high \ac{ATS} across a range of $p$, effectively preserving semantic information even under adverse conditions. 

Importantly, when we consider computational complexity—as detailed in Tab.~\ref{tab:complexity}, where alternative methods show exponential complexity growth with \(K\) or \(N\)—\ac{SemPA-Look} stands out by scaling linearly with \(K\). This balance of high \ac{ATS} and manageable computational complexity underscores its practicality.
Concretely, at $p=0.3$, \ac{SemPA-Look} outperforms Random \ac{PA} and transmission without packetization by 0.0306 and 0.0991, respectively, while being 0.0086, 0.0076, and 0.0056 lower than full search, \ac{SemPA-GBeam}, and \ac{GA}.

In Fig.~\ref{fig:plot_p_token}, subword-based tokenization is evaluated in the same setup in Fig.~\ref{fig:plot_p_word}, yielding slightly lower \ac{ATS} than word-based tokenization. In particular, \ac{SemPA-Look} experiences a marginal drop of 0.0061 when $p=0.3$ when tokenization is switched to subword-based tokenization. 

For the overall simulation results in \ref{sec:simulation_results}, subword-based tokenization yields a consistently lower \ac{ATS} than word-based tokenization under identical settings. This drop can be attributed to the fact that each important token is fragmented into multiple subwords: to reconstruct the critical word, all of its subwords must be received correctly. Losing even one subword prevents full reconstruction of the critical word, so on average recovering the complete token information becomes more difficult. As a result, the system becomes more vulnerable to semantic losses and the observed \ac{ATS} decreases.

\begin{table}[t!]
  \centering
  \caption{Computational Complexity for full search vs. GA vs. SemPA-GBeam vs. SemPA-Look.}
    \label{tab:complexity}
    \begin{tabular}{cc}
    \toprule
    \textbf{Method} & \textbf{Complexity} \\ 
    \midrule
    full search & $\displaystyle 2^K$ \\

    \ac{GA}, SemPA-GBeam & $\displaystyle G \cdot L \cdot 2^{N}$ \\

    SemPA-Look & $\displaystyle \left(N-1\right) \cdot P \cdot (k+1)$ \\
    \bottomrule
  \end{tabular}
\end{table}

\begin{figure}[t]
     \centering
     \subfloat[Word-based tokenization is applied.]{%
         \includegraphics[width=0.48\columnwidth]{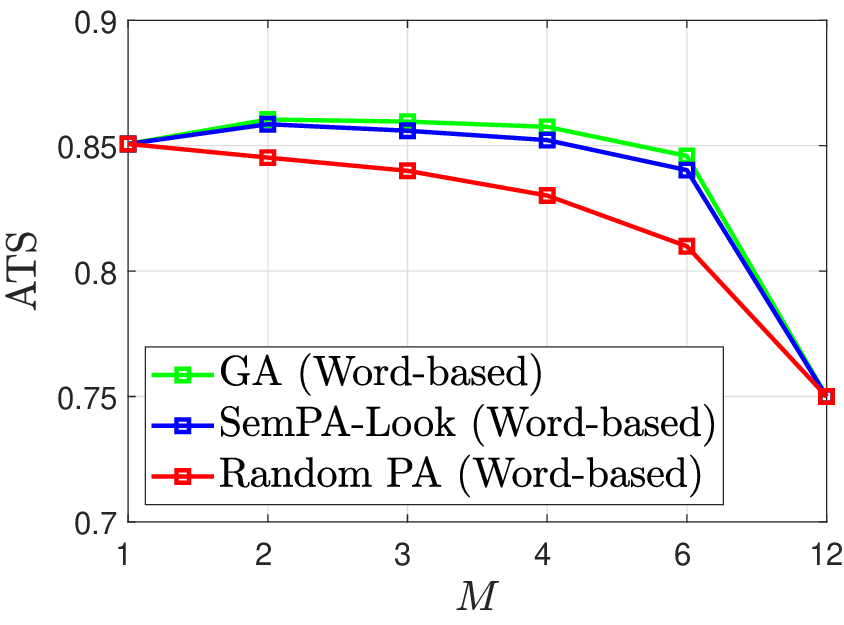}
         \label{fig:plot_M_word}
     }
     \hfill
     \subfloat[Subword-based tokenization is applied.]{%
         \includegraphics[width=0.48\columnwidth]{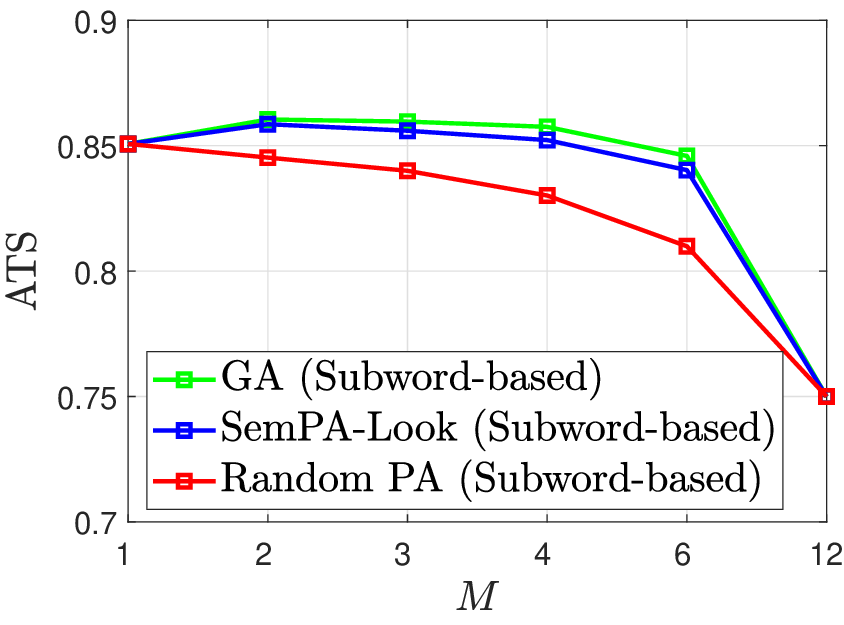}
         \label{fig:plot_M_token}
     }
     \caption{Impact of packet size \(M\) on the \ac{ATS}, highlighting the trade‑off between semantic loss and packet loss (\(K=12, P=10, k=4, p=0.25\)).}
     \label{fig:plot_M}
\end{figure}

\begin{figure}[t]
     \centering
     \subfloat[$p = 0.05$.]{%
         \includegraphics[width=0.48\columnwidth]{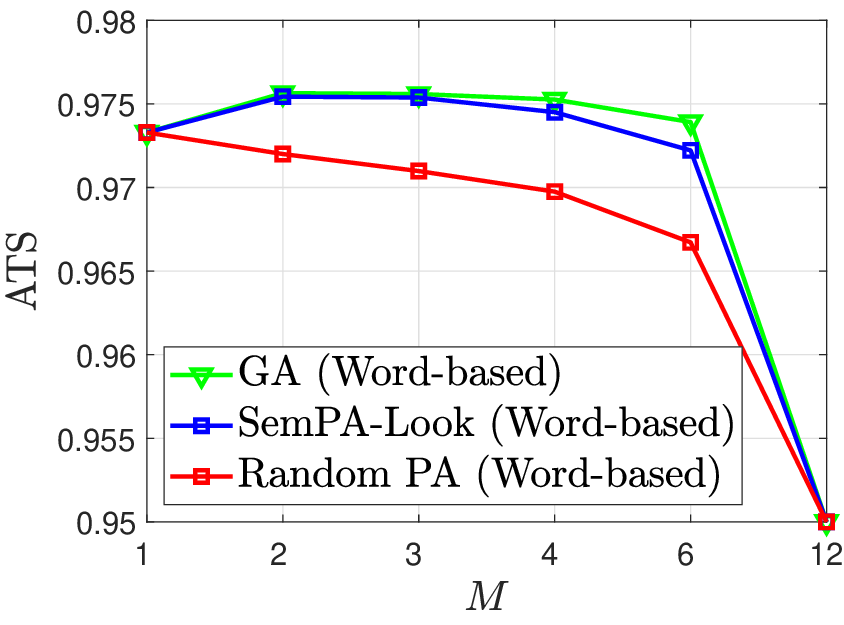}
         \label{fig:plot_M_low_p}
     }
     \hfill
     \subfloat[$p = 0.95$.]{%
         \includegraphics[width=0.48\columnwidth]{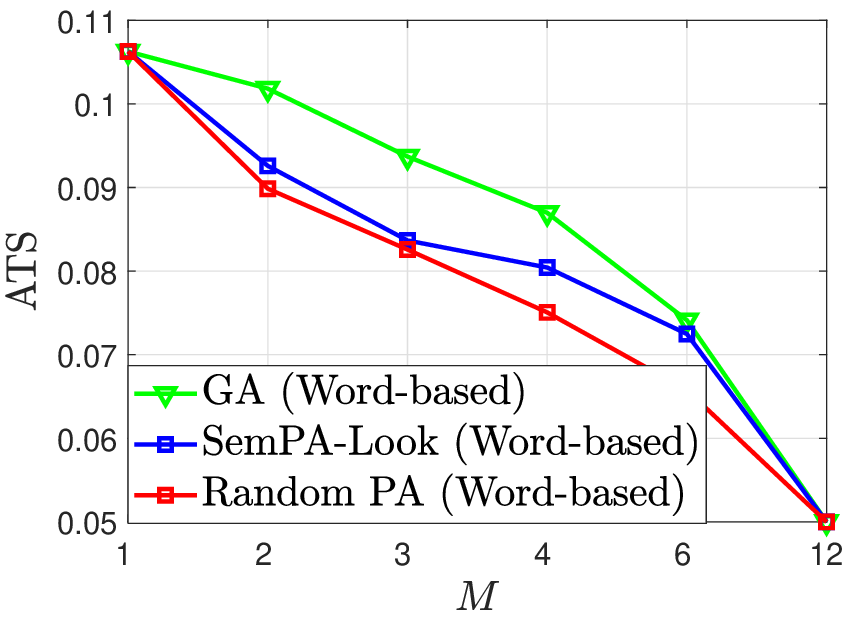}
         \label{fig:plot_M_high_p}
     }
     \caption{\ac{ATS} for different \(M\) values under extreme error conditions (\(p = 0.05\) and \(p = 0.95\)). Word-based tokenization is applied. The Figures show the trade-off between semantic loss and potential packet loss under different channel condition (\(K=12, P = 10, k = 4\)).} 
     \label{fig:plot_M_extreme_p}
\end{figure}


To analyze how packet size \(M\) affects ATS, we conduct experiments whose results are shown in Fig.~\ref{fig:plot_M} and Fig.~\ref{fig:plot_M_extreme_p}. In Fig.~\ref{fig:plot_M}, ATS shows a concave trend across different \(M\) values, peaking at \(M=2\) rather than \(M=1\).
Fig.~\ref{fig:plot_M_extreme_p} further examines this effect under extreme packet error rates (\(p=0.05\) and \(p=0.95\)). Under \(p=0.95\), \ac{ATS} is highest at \(M=1\), since the extreme loss probability makes larger packets prone to losing multiple tokens at once, favoring minimal packet lengths. However, when \(p=0.05\), similar to the results when $p=0.25$, the best performance at \(M=2\), rather than at \(M=1\) or \(M=12\), confirms the trade-off analyzed in Section~\ref{sec:trade_off}. In such scenarios, longer packets incur a larger semantic loss if they fail, while short packets (e.g., \(M=1\)) increase the overall packet loss rate. This trade-off between the semantic loss and the packet loss underlines why an intermediate packet length of \(M=2\) provides the optimal balance for high ATS.
In addition, as detailed in Tab.~\ref{tab:complexity}, reducing \(M\) also causes a steep increase in computational complexity. These findings highlight the necessity of selecting an appropriate \(M\) that optimally balances \ac{ATS} and computational complexity according to the channel error conditions.



\vspace{1em}
\noindent \textbf{\ac{ATS} Performance of \ac{PA} Methods on the Wiki-how Dataset}.\quad
We next evaluate our \ac{PA} method on the Wiki-how dataset, which comprises longer than MS-COCO. This extension to longer sentences poses significant computational complexity for full search, due to the exponential increase of complexity with respect to $K$. Hence, full search is excluded from evaluation. We present a comprehensive examination of how step size $k$, packet sizes $M$, population size $P$, and varying sentence length $K$ affect both \ac{ATS} and computational complexity.

\begin{figure}
\centering
\includegraphics[width=0.55\columnwidth]{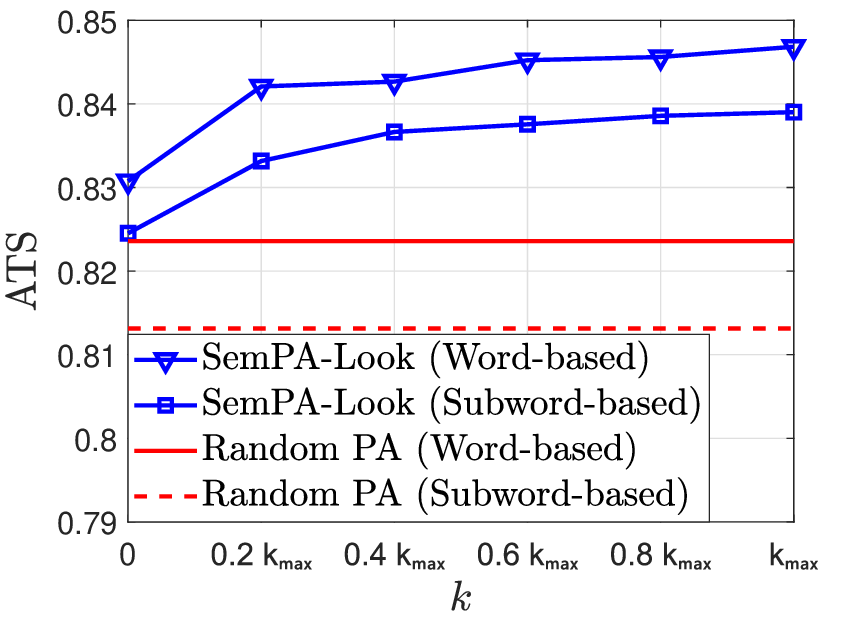}
\caption{\ac{ATS} as step size $k$ increases ($K=24, M=4, P = 10, p=0.25$). \ac{ATS} converges as $k$ increases.}\label{fig:plot_k}
\end{figure}

To evaluate the performance of \ac{SemPA-Look} and determine the optimal maximum step size \(k\) required for high \ac{ATS}, we investigate how \ac{ATS} varies as the step size \(k\) increases. 
In Fig.~\ref{fig:plot_k}, for \(K=24\) and \(M=4\), the maximum feasible step size at level 0 is \(k_{\mathrm{max}}=N-1=5\), decreasing by 1 with each subsequent level as the available token count diminishes. The results indicate that increasing \(k\) leads to higher ATS: by sampling more future packets and incorporating their \acp{RSS} into the decision for the current packet, the algorithm achieves more accurate \ac{ATS} prediction.
On the one hand, when \(K\) and \(N\) are large, increasing the maximum step size \(k\) allows sampling more future packets. However, our observations show that beyond a certain point, \ac{ATS} converges, which is the sufficient $k$ to predict \ac{ATS}.

\begin{figure}
     \centering
     \subfloat[\ac{ATS} as $P$ increases.]{%
         \includegraphics[width=0.48\columnwidth]{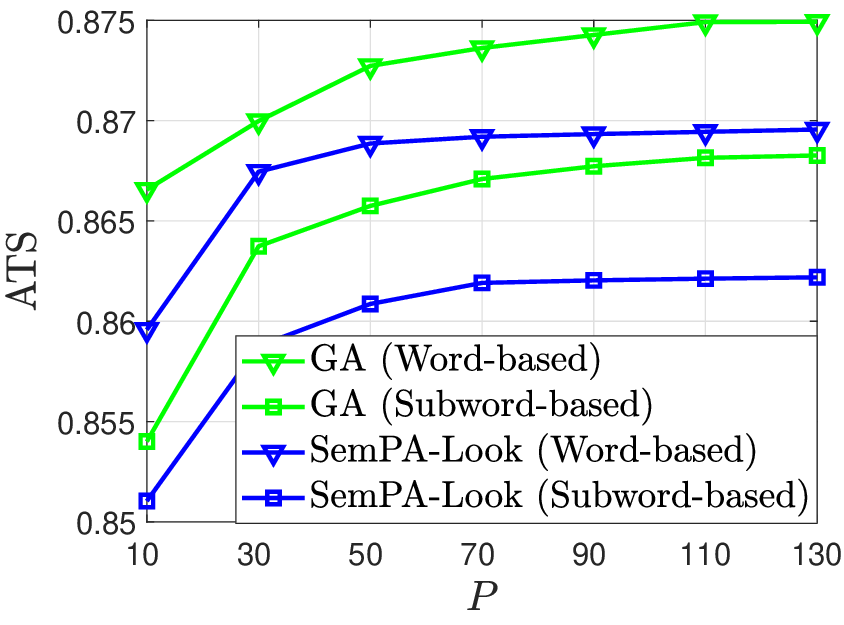}
         \label{fig:plot_P_score}
     }
     \hfill
     \subfloat[Computational complexity as $P$ increases.]{%
         \includegraphics[width=0.48\columnwidth]{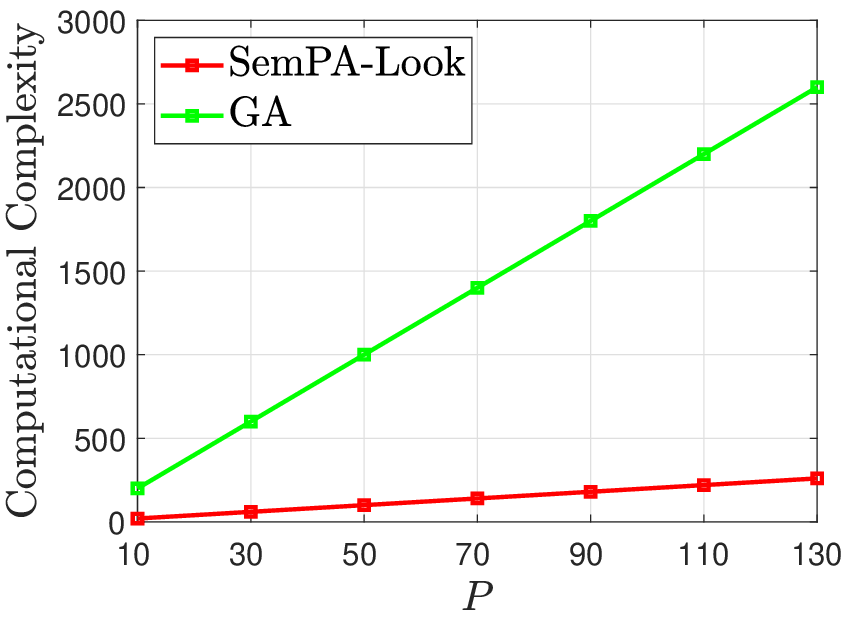}
         \label{fig:plot_P_complexity}
     }
     \caption{\ac{ATS} and computational complexity between original and decoded messages and computational complexity as $P$ increases ($K=60, M=30, G = 5, k = 4, p=0.25$).}

     \label{fig:plot_P}
\end{figure}

Fig.~\ref{fig:plot_P} compares the impact of the population size $P$ on the \ac{ATS} and computational complexity between \ac{GA} and \ac{SemPA-Look}. 
From Fig.~\ref{fig:plot_P_score}, it can be observed that as the $P$ increases from $P=10$ to $130$, both scores of \ac{GA} and \ac{SemPA-Look} increase. At $P=130$, \ac{GA} achieves a \ac{ATS} of $0.8749$ with word-based tokenization and $0.8683$ with subword-based tokenization. \ac{SemPA-Look} SemPA-Look obtains $0.8696$ for word-based tokenization, which is $0.0053$ lower than \ac{GA}, and $0.8622$ for subword-based tokenization, which is $0.0061$ lower than \ac{GA}.
However, in terms of computational complexity, as illustrated in Fig.~\ref{fig:plot_P_complexity}, \ac{GA} requires $2600$ text encoding steps at $P=130$, whereas \ac{SemPA-Look} demands only $260$ steps. This significant gap highlights the key advantage of \ac{SemPA-Look}: it attains nearly the same \ac{ATS} as \ac{GA} but with considerably fewer computations.



\begin{figure}
     \centering
     \subfloat[Word-based tokenization is applied.]{%
         \includegraphics[width=0.48\columnwidth]{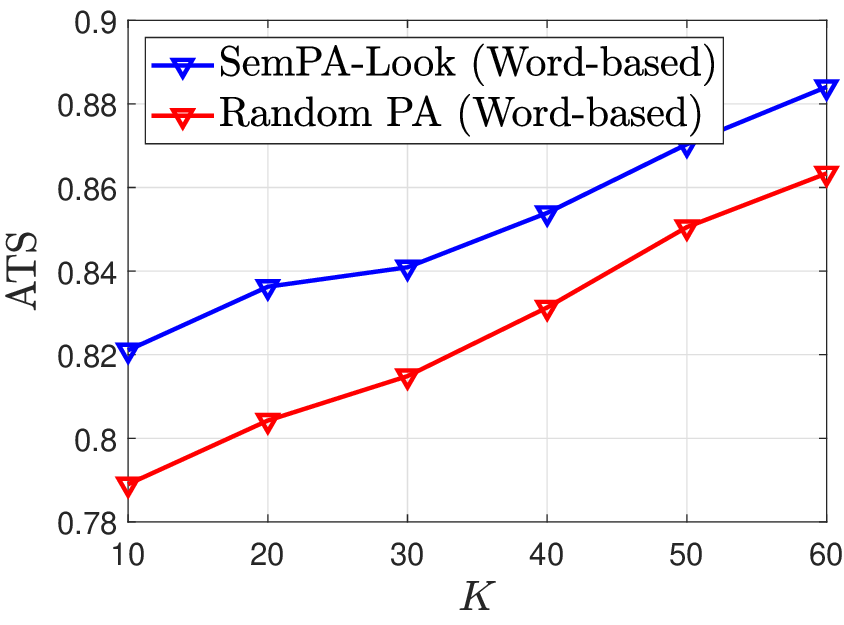}
         \label{fig:plot_K_word}
     }
     \hfill
     \subfloat[Subword-based tokenization is applied.]{%
         \includegraphics[width=0.48\columnwidth]{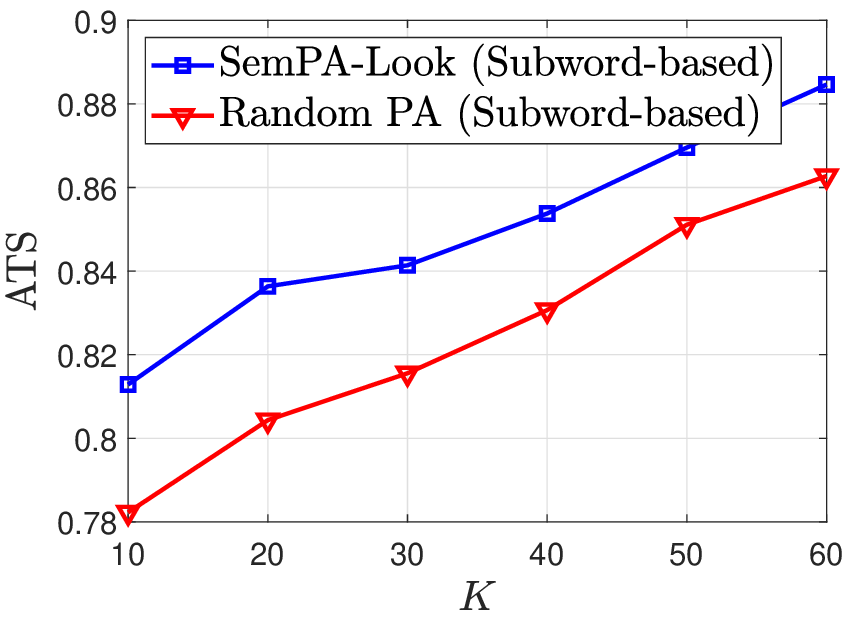}
         \label{fig:plot_K_token}
     }
     \caption{Impact of sentence length \(K\) on the \ac{ATS}, while $M$ is fixed to 5(\(k=4\), \(P=10\), \(p=0.25\)). Semantic loss reduction exceeds packet loss increase.}
     \label{fig:plot_K}
\end{figure}

Fig.~\ref{fig:plot_K} fixes \(M\) and varies \(K\) to examine the evolution of the packet loss–semantic loss trade‑off, complementing Fig.~\ref{fig:plot_M_extreme_p}, which analyzed packet size effects by varying \(M\) with \(K\) fixed. The results show that ATS increases with larger \(K\). This occurs because, although the packet loss increases linearly (since packet loss is  $\frac{K}{M}$), the semantic loss term \(-\sum_{\mathcal{C}\in \mathcal{G}}\phi\bigl(\mathcal{W}\setminus \mathcal{C},\mathcal{W}\bigr)\) decreases more rapidly because, as \(K\) increases, the set \(\mathcal{W}\setminus \mathcal{C}\) contains more tokens, resulting in a higher $\phi\bigl(\mathcal{W}\setminus \mathcal{C},\mathcal{W}\bigr)$ for each packet $\cC$. Moreover, the number of packets \(|\mathcal{G}| = \frac{K}{M}\) increases linearly with \(K\), which yields a proportionally greater reduction in the semantic loss term. Together, these effects result in an overall improvement in ATS, even as packet loss linearly increases with $K$. However, as the computational complexities of the PA methods also increase with \(K\), the improvement in \ac{ATS} comes at the expense of higher computational complexity, thus establishing a trade-off between \ac{ATS} and computational complexity.

\begin{figure}
     \centering
     \subfloat[Word-based tokenization is applied.]{%
         \includegraphics[width=0.48\columnwidth]{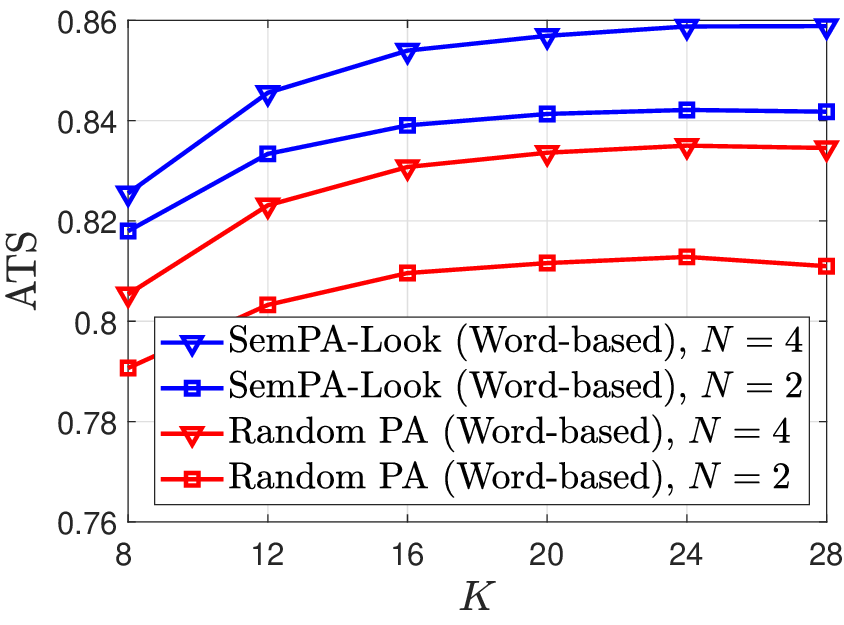}
         \label{fig:plot_M_over_K_word}
     }
     \hfill
     \subfloat[Subword-based tokenization is applied.]{%
         \includegraphics[width=0.48\columnwidth]{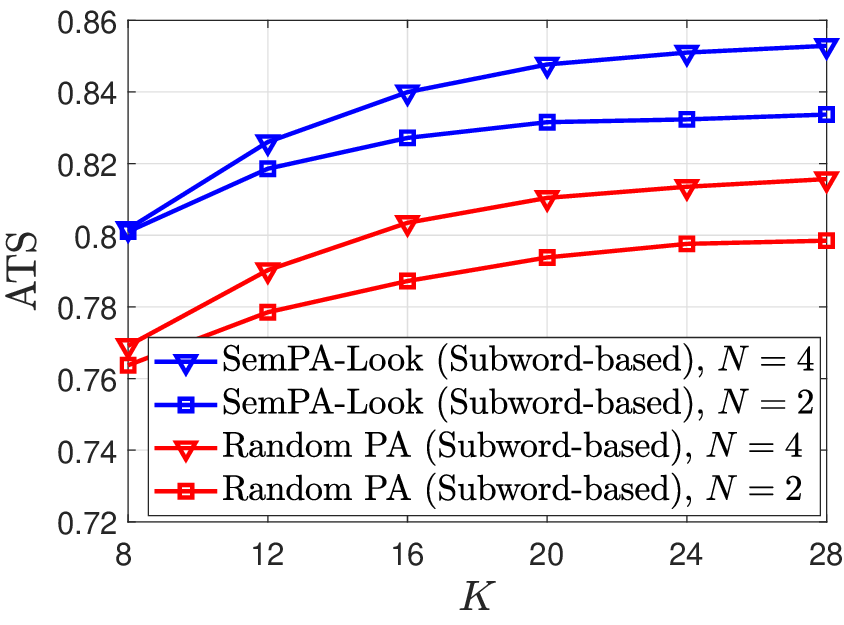}
         \label{fig:plot_M_over_K_token}
     }
     \caption{\ac{ATS} as \(K\) increases while $N$ is fixed at $2$ or $4$ (\(k=4\), \(P=10\), \(p=0.25\)). Under the fixed computational cost, ATS increases with \(K\) and eventually converges.} 
     \label{fig:plot_M_over_K}
\end{figure}

Noting that the complexity of \ac{SemPA-Look} increases linearly with $N = \frac{K}{M}$ and \ac{GA} grows exponentially with $N$, Fig.~\ref{fig:plot_M_over_K} shows the result when the ratio \(N\) is held constant (i.e., packet loss is constant) while both \(K\) and \(M\) increase. Unlike the result in Fig.~\ref{fig:plot_K} where the \ac{ATS} gradually increases as $K$ increases, it gradually converges as $K$ increases while $N$ is fixed. 
This observation leads to two key implications: (1) even with a constant \(N\), increasing both \(K\) and \(M\) results in a higher value of \(\phi\bigl(\mathcal{W}\setminus \mathcal{C},\mathcal{W}\bigr)\), thereby reducing semantic loss and improving ATS; and (2) these gains are bounded and eventually saturate as \(K\) grows with \(N\) fixed; fixing \(N\) keeps computational complexity constant, so increasing \(K\) does not indefinitely enhance ATS.

As a result, setting high $N$ is needed to obtain a high \ac{ATS} for large $K$. The complexity of the proposed \ac{SemPA-Look} and comparison methods is dependent on $N$, hence increasing $N$ increase the complexity of \ac{PA}. Note that the complexity of \ac{SemPA-Look} increases linearly with $N$, which is much lower than the exponential increase of full search and \ac{GA}, so the higher the $K$, the more effective \ac{SemPA-Look} is as a way to get a high \ac{ATS} with tolerable computational complexity.


\begin{figure}
     \centering
     \subfloat[Computational complexities as $K$ increases ($M = 2$).]{%
         \includegraphics[width=0.48\columnwidth]{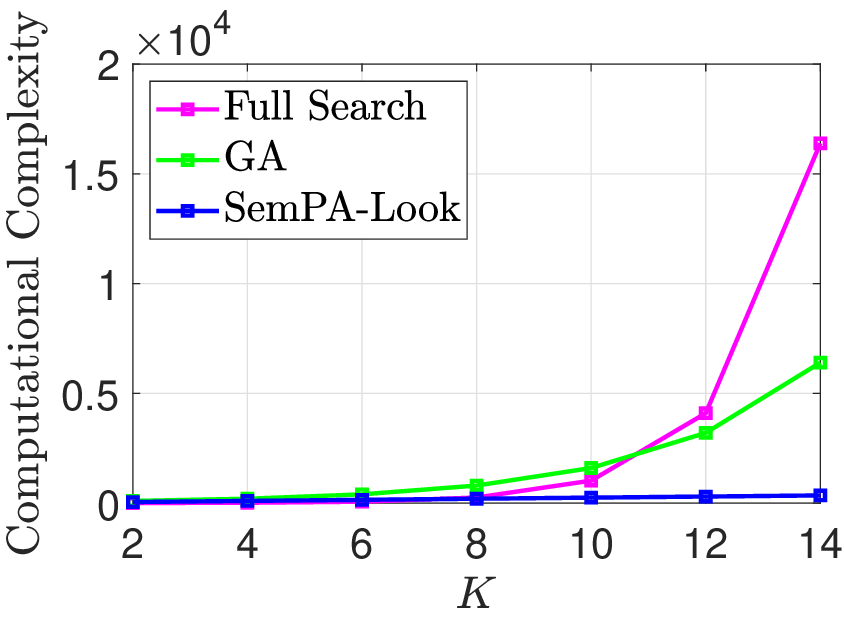}
         \label{fig:Complexity_analysis_K}
     }
     \hfill
     \subfloat[Computational complexities as $M$ increases ($K = 12$).]{%
         \includegraphics[width=0.48\columnwidth]{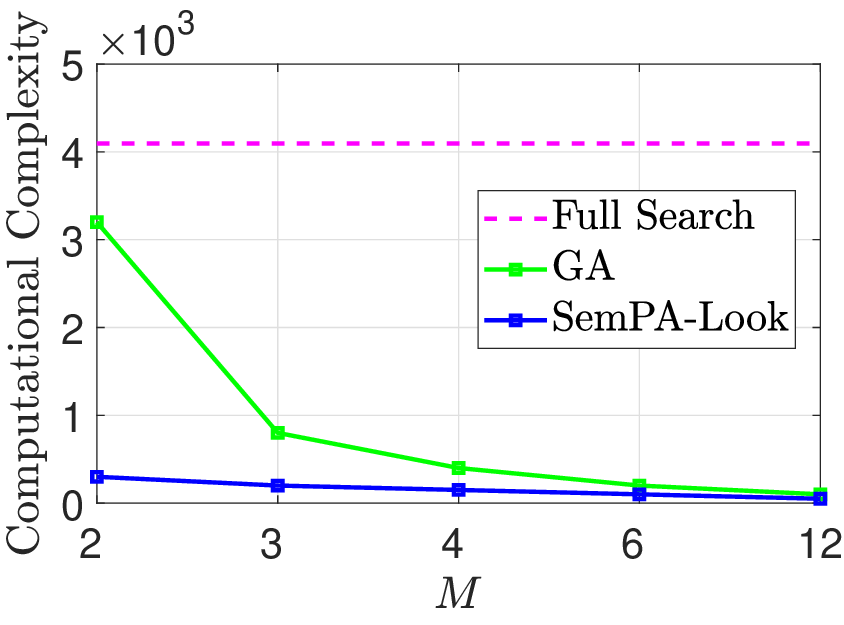}
         \label{fig:Complexity_analysis_M}
     }
     \caption{Computational complexities of different \ac{PA} methods ($k = 4, G = 5, P = 10, p = 0.25).$} 
     \label{fig:Complexity_analysis}
\end{figure}

Fig.~\ref{fig:Complexity_analysis} compares the computational complexity of different \ac{PA} methods. In Fig.~\ref{fig:Complexity_analysis_K}, \ac{SemPA-Look} at $K=14$ requires 18.3 times fewer computational complexity compared to \ac{GA} and 46.8 times fewer compared to full search, where $K=14$ is the moderate length of the message. The reason is that, as $K$ increases, other baseline methods encounter an exponential surge in complexity, whereas the proposed \ac{SemPA-Look} only increases linearly. In Fig.~\ref{fig:Complexity_analysis_M}, the computational complexity of \ac{SemPA-Look} at $M=2$ is 16 times fewer compared to \ac{GA}. The baseline \ac{PA} methods generate random sets of packets $\cG$ and directly measure their scores, thus incurring significantly higher overhead than the token-level optimization of \ac{SemPA-Look}.

\begin{figure*}[t]
\centering
\includegraphics[width=\textwidth]{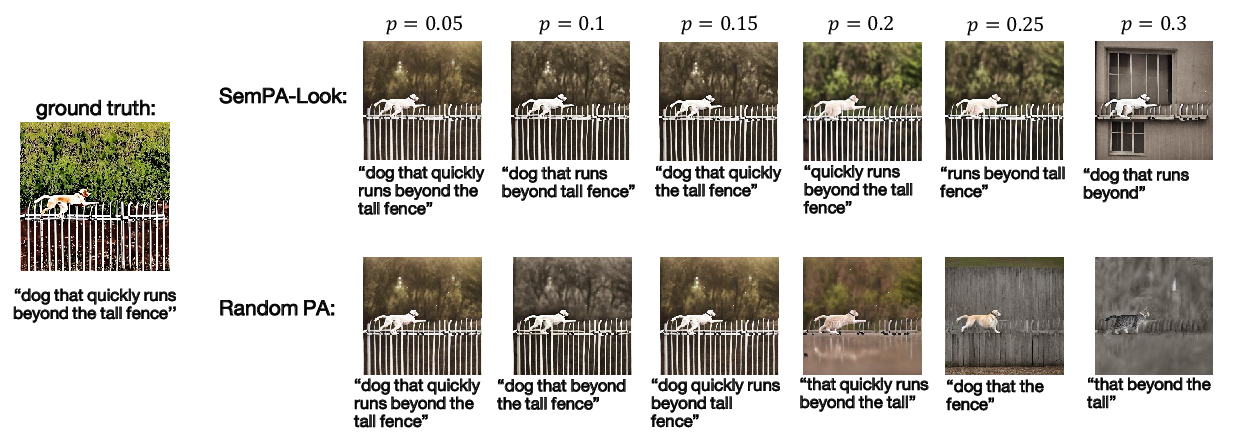}
\vspace{-1em}
\caption{Comparison of \ac{AIGC} of different \ac{PA} methods. Example AIGCs generated from captions reconstructed via \ac{SemPA-Look} and Random \ac{PA}, highlighting the impact of \ac{PA} on semantic preservation.}\label{fig:plot_AIGC}
\end{figure*}

\begin{figure}
     \centering
     \subfloat[Word-based tokenization is applied.]{%
         \includegraphics[width=0.48\columnwidth]{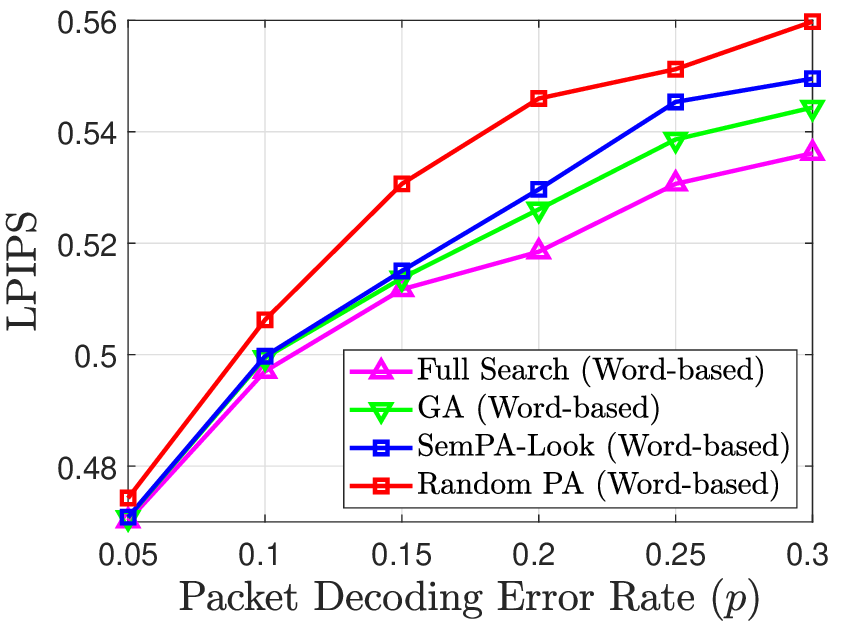}
         \label{fig:plot_word_LPIPS}
     }
     \hfill
     \subfloat[Subword-based tokenization is applied.]{%
         \includegraphics[width=0.48\columnwidth]{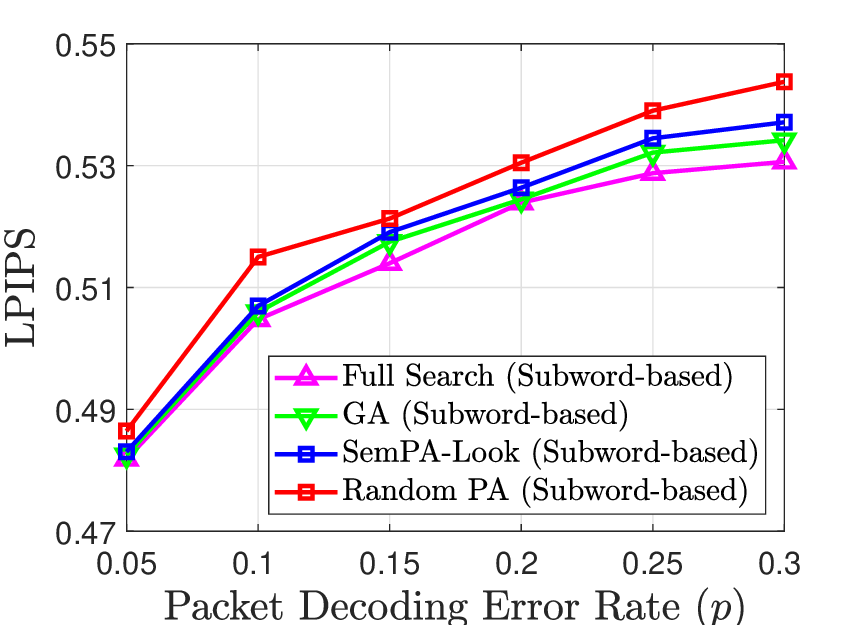}
         \label{fig:plot_token_LPIPS}
     }
     \caption{Average LPIPS between between images generated from received messages and original messages across different \ac{PA} methods as $p$ increases ($K = 8, M = 4, k = 4, P = 10$).} 
     \label{fig:plot_LPIPS}
\end{figure}

\vspace{1em}
\noindent \textbf{AI-Generated Image Quality from received captions using \ac{PA} Methods on MS-COCO Dataset}.\quad
Finally, for the image-level performance analysis, we conduct experiments using the MS-COCO image-caption dataset to evaluate various \ac{PA} methods. In our setup, image–caption pairs are first extracted from the dataset. Next, a coarse version of the original image, denoted by \(\cI_C\), is generated using Canny edge detection, and this coarse image is used as a conditioning input to a Stable Diffusion model \cite{rombach2022} at \ac{RX}.

At \ac{RX}, the caption information is reconstructed from the received packet group, denoted by $\hat{\mathcal{G}}$, using various \ac{PA} methods. The reconstructed caption, $\widehat{\cW}$, together with the coarse image, is fed into the diffusion model to generate the final image:
\begin{equation}
    \cI_R = S\bigl(\cI_C,\, \hat{\mathcal{G}}\bigr),
\end{equation}
where $S(\cdot,\cdot)$ represents the image synthesis process via Stable Diffusion, with $\cI_C$ serving as the visual guide and $\hat{\mathcal{G}}$ encoding the semantic information from the transmitted caption.

The fidelity of the reconstructed image \(\cI_R\) relative to the original \(\cI\) is measured using the Learned Perceptual Image Patch Similarity (LPIPS) metric:
\begin{align}
    \text{LPIPS}(\cI_R, \cI) = \sum_{l} \frac{1}{H_l W_l} \sum_{i,j} \left\| f_{l}(\cI_R) - f_{l}(\cI) \right\|_{2}^{2},
\end{align}
where \(f_{l}(\cdot)\) represents the feature map extracted from the \(l\)th layer of a pre-trained AlexNet, with \(H_l\) and \(W_l\) denoting the height and width of that feature map, respectively. A lower LPIPS score indicates a higher perceptual similarity between the images.

Fig.~\ref{fig:plot_AIGC} presents example AIGC outputs generated from the reconstructed captions, while Fig.~\ref{fig:plot_LPIPS} shows the average LPIPS scores of generated \ac{AIGC} as \(p\) increases. Overall, \ac{SemPA-Look} achieves LPIPS scores that are competitive with Full Search and \ac{GA}, and lower than those of Random \ac{PA}. 
In particular, the trend observed in Fig.~\ref{fig:plot_p} confirms that as the semantic content of the caption increases, the quality of the generated images improves, which validates the effectiveness of \ac{SemPA-Look}.

For example, in Fig.~\ref{fig:plot_AIGC}, with the original caption ``dog that quickly runs beyond the tall fence'' and \(p=0.3\), \ac{SemPA-Look} reconstructs it as ``dog that runs beyond'', thereby preserving the essential meaning that a dog leaping over an obstacle. In contrast, Random \ac{PA} reconstructed the caption as ``that beyond the tall'', which not only omits the reference to a dog but also yields an image showing a cat without fence, thus significantly distorting the intended meaning.

For example, in Fig.~\ref{fig:plot_AIGC}, with the original caption ``dog that quickly runs beyond the tall fence'' and \(p=0.3\), \ac{SemPA-Look} reconstructs it as ``dog that runs beyond'', thereby preserving the essential meaning of a dog leaping over an obstacle. In contrast, Random \ac{PA} loses the packet carrying the core semantic content and reconstructs the caption as ``that beyond the tall'', omitting the dog entirely and producing an image of a cat without a fence, where the intended meaning is severely distorted.



\section{Conclusion}\label{sec:conclusion}
In this paper, we proposed low-complexity \ac{PA} for text-based semantic communication that significantly improves the preservation of critical semantics over unreliable channels. Our approach introduces a novel \ac{RSS} that quantitatively captures the semantic importance of packets and employs \ac{SemPA-Look} to explicitly consider the inter-packet dependencies arising from sequential transmission. Our method effectively determines the optimal number of packet transmissions and reduces semantic loss. As a result, the overall semantic fidelity of the reconstructed message, as measured by cosine similarity between the original and received text, is substantially enhanced with low computational complexity.

Extensive simulation results demonstrate that our proposed method outperforms conventional \ac{PA} methods. Our approach achieves higher token similarity while reducing computational complexity compared to exhaustive search and genetic algorithm-based methods. The experiments reveal that our method is robust across a variety of channel conditions and scales efficiently with increasing message length.

Future research will focus on further refining packet aggregation methods by exploring advanced tokenizations, such as those employed in large foundation models, to better preserve key semantic information under dynamic channel environments. In addition, extending our framework to support multi-modal semantic communication and developing real-time optimization techniques for practical wireless systems are promising directions that could further improve the efficiency and robustness of next-generation semantic communication networks.

\bibliographystyle{IEEEtran}
\bibliography{si}
\end{document}